\documentclass[twocolumn]{aastex}
\usepackage{amsmath}
\usepackage[figuresright]{rotating}
\graphicspath{{figures/}}
\usepackage{subfigure}
\usepackage{float}
\usepackage[]{color}
\definecolor{ForestGreen}{RGB}{34,139,34}

\begin{document}
%
%
%%%%%%%%%%%%%%%%%%%%%%%%%%%%%%%%%%%%%%%%%%%%%%%%%%%%%%%%%%%%%%%%%%%%%%%%%%%%%%%%%%%%%%%%%%%%%%%%
%
%
\title{Self-Consistent Heating of the Magnetically Closed Solar Corona:

Generation of Nanoflares, Thermodynamic Response of the Plasma and Observational Signatures
}
%
%
%%%%%%%%%%%%%%%%%%%%%%%%%%%%%%%%%%%%%%%%%%%%%%%%%%%%%%%%%%%%%%%%%%%%%%%%%%%%%%%%%%%%%%%%%%%%%%%
%
%
\author[0000-0003-4023-9887]{Craig D. Johnston}
\affiliation{Department of Physics and Astronomy,
George Mason University,
Fairfax, VA 22030, USA}
\affiliation{Heliophysics Science Division,
NASA Goddard Space Flight Center,
Greenbelt, MD 20771, USA}
\correspondingauthor{Craig D. Johnston}
\email{craig.d.johnston@nasa.gov, cjohn44@gmu.edu}

\author[0000-0002-1198-5138]{Lars K. S. Daldorff}
\affiliation{Department of Physics,
Catholic University of America,
Washington, DC 20064, USA}
\affiliation{Heliophysics Science Division,
NASA Goddard Space Flight Center,
Greenbelt, MD 20771, USA}

\author[0000-0003-2255-0305]{James A. Klimchuk}
\affiliation{Heliophysics Science Division,
NASA Goddard Space Flight Center,
Greenbelt, MD 20771, USA}

\author[0000-0003-4225-8520]{Shanwlee Sow Mondal}
\affiliation{Department of Physics,
Catholic University of America,
Washington, DC 20064, USA}
\affiliation{Heliophysics Science Division,
NASA Goddard Space Flight Center,
Greenbelt, MD 20771, USA}

\author[0000-0001-9642-6089]{Will T. Barnes}
\affiliation{Department of Physics,
American University,
Washington, DC 20016, USA}
\affiliation{Heliophysics Science Division,
NASA Goddard Space Flight Center,
Greenbelt, MD 20771, USA}

\author[0000-0003-0072-4634]{James E. Leake}
\affiliation{Heliophysics Science Division,
NASA Goddard Space Flight Center,
Greenbelt, MD 20771, USA}

\author[0000-0002-6391-346X]{Jack Reid}
\affiliation{School of Mathematics and Statistics, 
University of St Andrews,  
St Andrews,  KY16 9SS, UK}

\author[0000-0001-8732-8284]{Jacob D. Parker}
\affiliation{Heliophysics Science Division,
NASA Goddard Space Flight Center,
Greenbelt, MD 20771, USA}

%
%
%%%%%%%%%%%%%%%%%%%%%%%%%%%%%%%%%%%%%%%%%%%%%%%%%%%%%%%%%%%%%%%%%%%%%%%%%%%%%%%%%%%%%%%%%%%%%%%%
%
%
\begin{abstract}

The energy that heats the magnetically closed solar corona originates in the complex motions of the massive photosphere. 
Turbulent photospheric convection slowly displaces the footpoints of coronal field lines, causing them to become twisted and tangled. 
Magnetic stresses gradually build until reaching a breaking point when the field reconnects and releases a sudden burst of energy. 
We simulate this basic picture of nanoflares using a high-fidelity, three-dimensional, multi-stranded magnetohydrodynamic simulation that starts with a fully stratified atmosphere. 
This simulation includes the effects of field-aligned thermal conduction and optically thin radiation and uses the state-of-the-art Transition Region Adaptive Conduction (TRAC) method to capture the response of the plasma to the nanoflare heating. 
We find that our physical model supports a unified explanation for both the diffuse emission observed in active regions and the bright coronal loops. 
Specifically, our results suggest that the diffuse emission originates from spatially and temporally uncorrelated nanoflares, whereas coherent clusters of nanoflares -- nanoflare storms -- are responsible for the formation of bright coronal loops. 
Quantitative comparisons between the simulated emission and observed characteristics of coronal loops show that key observed properties -- such as loop widths, lifetimes and cross sections -- are reasonably well reproduced by the model. 
The idea that avalanche spread naturally leads to circular cross sections in coronal loops is strongly supported.
Our results also suggest that phase differences in heating and cooling events across neighboring magnetic flux strands are a plausible explanation for the anomalous cross-field motions of coronal loops that were recently reported in high-resolution observations.

\end{abstract}
%
%
%%%%%%%%%%%%%%%%%%%%%%%%%%%%%%%%%%%%%%%%%%%%%%%%%%%%%%%%%%%%%%%%%%%%%%%%%%%%%%%%%%%%%%%%%%%%%%%%
%
%
% The AAS Journals now uses Unified Astronomy Thesaurus concepts:
% https://astrothesaurus.org
%
%
%
%
%%%%%%%%%%%%%%%%%%%%%%%%%%%%%%%%%%%%%%%%%%%%%%%%%%%%%%%%%%%%%%%%%%%%%%%%%%%%%%%%%%%%%%%%%%%%%%%
%
%
\section{Introduction} \label{sec:intro}

Solar active regions have tremendous complexity on small spatial scales \citep[see e.g.,][]{Reale2014}.
The magnetic field is subdivided into an enormous number of quasi-independent magnetic flux strands -- of order $10^5$ in a single active region \citep{Klimchuk2015}.
Rooted in the photosphere, these coronal flux strands are twisted and tangled by turbulent convective motions.
Electric current sheets form at the boundaries between the strands.
This gradual buildup of magnetic energy culminates in impulsive reconnection once critical conditions are met, releasing a burst of energy --  a nanoflare \citep{Parker1983, Parker1988} -- that heats the coronal plasma.
Each nanoflare is too small to be detected individually – though mission concepts are being developed to one day do so \citep[see e.g.,][]{Rabin2023} – but it is proposed that they collectively heat the corona to its multi-million degree temperatures. The forthcoming Multi-slit Solar Explorer (MUSE) spectrometer \citep[e.g.,][]{DePontieu2022} will also investigate coronal heating with high spatial, temporal, and spectral resolution.

Using magnetohydrodynamic (MHD) models to study magnetically closed loops in the
solar atmosphere, we have learned a great deal about nanoflare reconnection in the corona \citep[see e.g.,][]{Gudiksen2005,Bingert2011,Hansteen2015,Rempel2017}, including fundamental properties about reconnection onset \citep{Leake2020,Klimchuk2023b,Leake2024}.
Simulating the plasma response to the heating in such models requires a physical connection between the corona, transition region (TR), and chromosphere to properly account for the mass and energy exchange that occurs through field-aligned thermal conduction and flows (evaporation and draining).
In particular, to make meaningful comparisons with observations, the narrow, computationally demanding TR must be accurately treated throughout all phases of an impulsive heating event \citep{Bradshaw2013,Johnston2017a,Johnston2017b,Johnston2019b}

Existing MHD simulations of active regions do not include the small-scale complexity of
the magnetic field, and many do not properly treat the TR \citep[e.g.,][]{Rempel2017,ChenRempel2022,Lu2024}. 
The number of numerical grid cells required to capture the $10^5$ current sheets is prohibitive, and resolving the narrow TR across an entire active region is an extreme challenge. 
Consequently, the heating mechanism in the simulations is different from that on the real Sun. 
It typically involves artificially large Ohmic or numerical dissipation of large-scale currents, which is far more gradual than the impulsive reconnection of nanoflares.

These challenges have led to the development of multi-strand MHD simulations that simulate only a small fraction of the magnetic flux \citep[see e.g.,][]{Hood2016,Pontin2017,Knizhnik2018,Reid2018, Howson2022,Breu2022,Cozzo2023}, instead of trying to treat an entire active region in a single simulation.
Such simulations limit the number of magnetic flux strands and current sheets to improve resolution, but retain a sufficient number so that collective behavior can still occur.
Each multi-strand simulation, characterized by one unique magnetic field strength and loop length, then represents one building block of an entire active region.

Recently, using a multi-strand MHD simulation, \cite{Klimchuk2023a} found that individual magnetic flux strands experience short-term brightenings, both scattered throughout the computational volume and in localized clusters. 
They associated the former with the diffuse component of the observed corona and the latter with distinct coronal loops. 
Coronal loops had previously been attributed to \lq\lq storms\rq\rq\ of nanoflares based on observational considerations \citep[e.g.,][]{Klimchuk2009}, and this provided a physical explanation for their origin, including their roughly circular cross section \citep[e.g.,][]{Klimchuk2020}.

The results of \cite{Klimchuk2023a} are limited, however, by several simplifying assumptions. 
Most significant is their treatment of the plasma response to the impulsive heating and subsequent production of emission. 
The MHD simulation did not include the effects of field-aligned thermal conduction and optically thin radiation, or a lower atmosphere (TR and chromosphere). 
The crucially important processes of conductive and radiative cooling, and plasma exchange between the corona and lower atmosphere were treated with a highly simplified \lq\lq cooling model\rq\rq\ that is based on results from zero dimensional \lq\lq Enthalpy-based Thermal Evolution of Loops\rq\rq\ \citep[EBTEL;][]{Klimchuk2008,Cargill2012a,Cargill2012b} simulations. 
This cooling model was applied to the MHD simulation output post facto.
Thus, there was also no coupling between the magnetic and \lq\lq thermodynamic\rq\rq\ processes.
Furthermore, the photospheric driving consisted of rotational cells at fixed locations. 
There was no translational component to the driving.
\cite{Klimchuk2023a} argued that, because of differences in size scale, the roughly circular loop cross sections are not an artifact of the purely rotational driving.
However, this is not well established. 
Finally, the aspect ratio of the system – the ratio of loop length to both driver scale and loop diameter – was much smaller than on the Sun.

In the simulation presented in this paper, we address all these shortcomings by incorporating the following improvements into our model:
(1)~a~proper thermodynamic treatment of the atmosphere with thermal conduction and optically thin radiation included using the MHD Transition Region Adaptive Conduction (TRAC) method \citep{Johnston2020,Johnston2021}, 
(2)~a~more~random form of photospheric driving that includes both rotational and translational motion, and
(3)~a more~realistic loop length relative to the characteristic size of the driver flows.
The outcome is a rigorous, self-consistent simulation of nanoflare  heating in the magnetically closed solar corona that supports a unified explanation for both the diffuse emission observed in active regions and the bright coronal loops \citep{Klimchuk2009,Klimchuk2015}.
In particular, consistent with \cite{Klimchuk2023a}, we find that randomly scattered nanoflares can form the diffuse emission, while storms of collective nanoflares can produce the coronal loops.
Furthermore, our simulation reveals that circular cross sections of coronal loops naturally arise from avalanche spread during nanoflare storms.

A complete description of our numerical model is provided in Section \ref{sec:model}.
Section \ref{sec:results} presents our results on the generation of nanoflares, the response of the plasma and quantitative comparisons between the simulated emission and observed characteristics of coronal loops.
Emphasis is placed on extreme ultraviolet (EUV) emission produced in the temperature-dependent channels of the Atmospheric Imaging Assembly \citep[AIA;][]{Lemen2012} onboard the Solar Dynamics Observatory \citep[SDO;][]{Pesnell2012}.
Finally, in Section \ref{sec:discuss}, we summarize our findings, compare them with previous simulations, and discuss their implications for recent high-resolution observations.
%
%
%%%%%%%%%%%%%%%%%%%%%%%%%%%%%%%%%%%%%%%%%%%%%%%%%%%%%%%%%%%%%%%%%%%%%%%%%%%%%%%%%%%%%%%%%%%%%%%
%
%
\section{Numerical Model} \label{sec:model}

\subsection{Governing Equations}

To model the generation of nanoflares and response of the plasma in the magnetically closed solar corona the following set of MHD equations are solved numerically using version 3.3 of the Lagrangian Remap (LaRe) code \citep{Arber2001}:
\begin{align}
    &
    \frac{\partial\rho}{\partial t}  
    + \nabla \cdot (\rho {\bf v})
    = 0; 
    \label{Eqn:mhd_continuity}
    \\[1mm]
    &
    \rho \frac{D{\bf v}}{Dt}
    = - 
    \nabla P - \rho {\bf g} + {\bf j \times B} + 
    {\bf F}_{\textrm{shock}};
    \label{Eqn:mhd_motion}
    \\[1mm]
    &
    \frac{\partial {\bf B}}{\partial t}
    = 
    \nabla \times ({\bf v \times B}); %- \nabla \times (\eta {\bf j});
    \label{Eqn:mhd_induct}
    \\[1.5mm]
    &
    \rho \frac{D \epsilon}{Dt}
    = -P\nabla \cdot{\bf v} 
    + Q_{\textrm{shock}} \!
    -\nabla \cdot {\bf q}  
    -  n^2 \Lambda(T);
    \label{Eqn:mhd_ee}
    \\[1.5mm]
    &
    P = 2 {\rm{k_B}} n T.
    \label{Eqn:mhd_gas_law}
\end{align} 
In these equations, $\rho$ is the mass density,  
${\bf v}$ is the velocity, 
$P$ is the gas pressure of a fully ionized plasma, 
${\bf g}$ 
is the gravitational acceleration,
${\bf j=(\nabla \times B})/\mu_0$ is the current density and
${\bf B}$ is the magnetic field.
The specific 
internal energy density is given by $\epsilon=P/(\gamma-1) \rho $
(where $\gamma=5/3$ is the ratio of specific heats), 
$n$ is the electron number density ($n=\rho/1.2\, \rm{m_p}$, where $\rm{m_p}$ is the
proton mass),
$\rm{k_B}$ is the Boltzmann constant and
$T$ is the temperature.

${\bf F}_{\textrm{shock}}$ in the momentum Eq.~\eqref{Eqn:mhd_motion} represents a viscous force that is finite at shocks but reduces to zero for smooth flows. 
At shocks, the shock jump conditions are satisfied through the use of shock viscosities \citep{Arber2001}.
This allows heating due to shocks to be captured as a viscous heating $Q_{\textrm{shock}}$ which is added into the thermal energy Eq. \eqref{Eqn:mhd_ee}.
Such approach has been extensively used in LaRe to simulate the heating from magnetic reconnection in the solar corona \citep{Hood2016,Reid2018,Reid2020,Leake2020,Threlfall2021,Howson2022,Leake2024}. 

However, in contrast to these previous simulations, we do not include an explicit resistivity $\eta$ in the induction Eq.~\eqref{Eqn:mhd_induct} or corresponding Ohmic heating term $\eta {\bf j}^2$ in Eq.~\eqref{Eqn:mhd_ee}.
This is to avoid unrealistic heating in places where it would not occur without artificially large resistivity. 
Instead, the field line breaking that is required for magnetic reconnection is provided by numerical resistivity in our simulation and occurs where very steep magnetic gradients develop. 
This allows us to capture impulsive energy release events at narrow current sheets -- nanoflares -- and the conversion of energy as it occurs for nanoflares in the solar corona: from magnetic to kinetic and then into thermal energy through viscous dissipation of the reconnection outflow jets \citep[e.g.,][]{Bareford2015,Reid2020}.

Thus, we take the view that viscous shock heating dominates any actual Ohmic heating that occurs in the real corona. During reconnection, most of the input magnetic energy flux is converted into kinetic energy and enthalpy fluxes of the exhaust, with only a small fraction going into direct Ohmic dissipation. The exhaust energy is then thermalized at shocks. This energy budget is supported by in situ observations of reconnection in the solar wind \citep{Mistry2017} and magnetopause \citep{Phan2014}.

We include field-aligned thermal conduction and optically thin radiation in the thermal energy Eq.~\eqref{Eqn:mhd_ee} to model the enthalpy exchange between the corona and TR in response to these impulsive heating events \citep{Johnston2019a}.
In particular, the heat flux vector ${\bf q}$ is based on the \cite{Braginskii1965} formulation  in the presence of a magnetic field, which recovers the field-aligned Spitzer-H{\"a}rm parallel thermal conductivity in the strong field limit \cite[]{Spitzer1953}.  
The radiative loss function $\Lambda(T)$ of an optically thin plasma is approximated using the 
piecewise continuous function defined in
\cite{Klimchuk2008}.

The TRAC method developed by \cite{Johnston2020,Johnston2021} accurately captures the thermodynamic coupling between the corona, TR and chromosphere without the need for high spatial resolution.  In this paper, we used the localized MHD formulation of TRAC presented in \cite{Johnston2021} because of its unique ability to (1)~accurately treat the TR throughout all phases of an impulsive heating event and (2)~automatically account for changes in magnetic field line connectivity. 
This formulation thus facilitated simultaneous modeling of the coronal magnetic field and thermodynamic response of the stratified plasma.

\subsection{Initial Conditions \& Boundary Conditions}

The simulation is initialized with a $10^{-2}$~T ($100$~G) uniform magnetic field in the $z$ direction of a three dimensional (3D) computational domain of extents 
$X \times Y \times  Z = [0, 33.75]~\textrm{Mm} \times [0, 33.75]~\textrm{Mm} \times [-15, 115]~\textrm{Mm}$.
This domain corresponds to a small subset of the magnetic flux in an active region.
A uniform numerical grid comprised of $256 \times 256 \times 512$ points is used to resolve the domain, giving a spatial resolution of approximately 130~km in the horizontal directions and 250~km in the vertical direction.

We model a coronal arcade with a Parker geometry \citep{Parker1972} by using a spatially varying gravity in $z$ which corresponds to that of a semi-circular arcade,
\begin{align}
    {\bf g} = g_\odot 
    \cos    
    \left( 
    \frac{\pi (z-z_b)}{L_z}
    \right)
    {\bf \hat {z}}.
    \label{Eqn:grav}
\end{align} 
Here, $g_\odot = 274~\textrm{ms}^{-2}$, $z_b=-15$~Mm and $L_z=130$~Mm so that gravity reverses sign about the $Z$-midplane ($z=50$~Mm). 
\begin{figure}
    \hspace{-4mm}
    \includegraphics[width=0.5\textwidth]{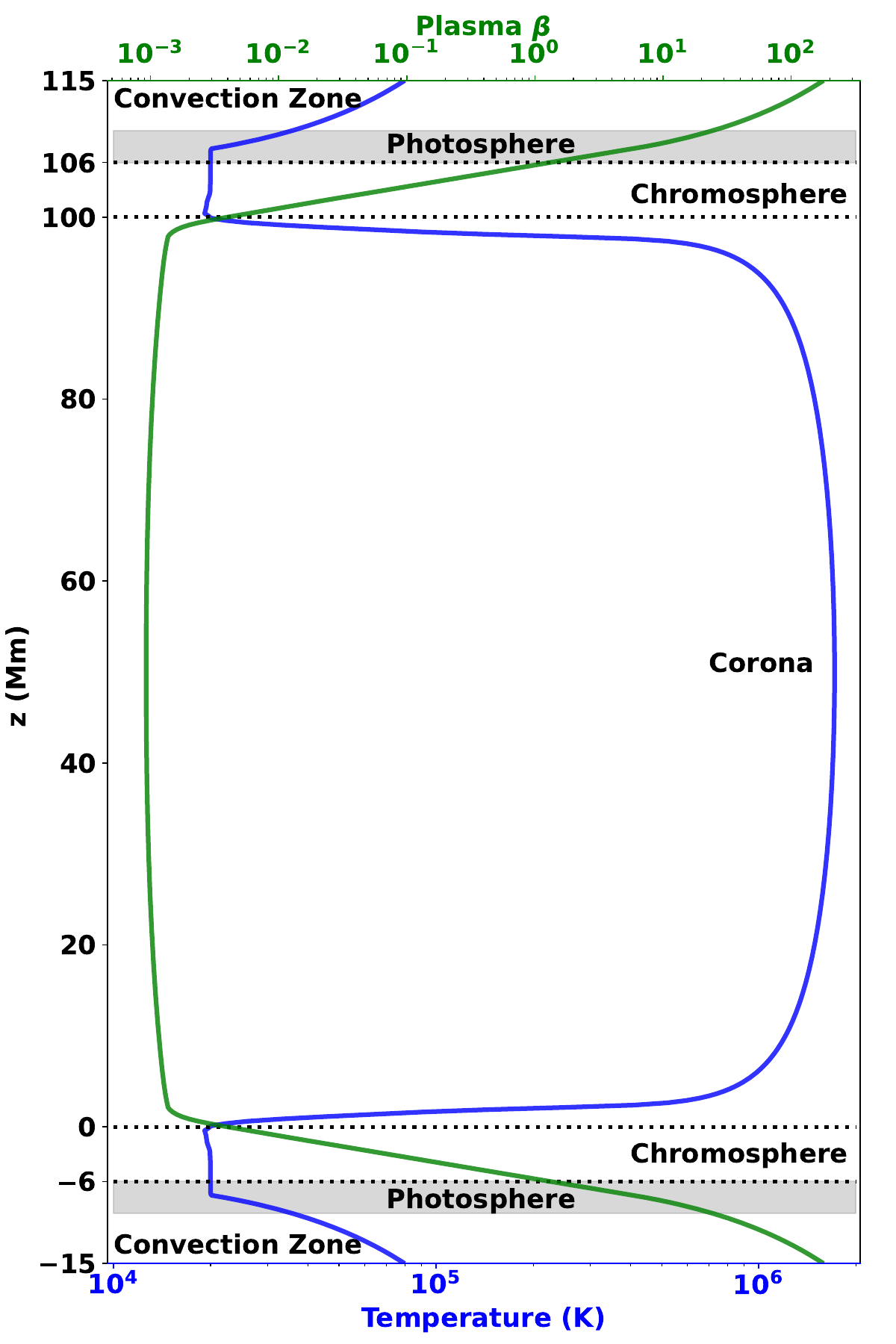}
    \vspace{-7mm}
    \caption{Plasma $\beta$ (top axis) and temperature (bottom axis) as functions of height for the initial hydrostatic atmosphere. 
    The shaded regions indicate the driving layers that are used to control the imposed photospheric motions.
    }
  \label{fig:hydro_atmosphere} 
\end{figure}

Following \cite{Leake2022} and \cite{Johnston2025}, the initial thermodynamic state is stratified from the bottom boundary upwards using a hydrostatic atmosphere consisting of a convection zone ($z < -6$~Mm), photosphere ($z = -6$~Mm), chromosphere ($-6 < z < 0$~Mm), TR ($z=0$~Mm) and corona ($ 0 < z < 50$~Mm).
This thermally structured atmosphere is then mirrored about the $Z$-midplane ($z=50$~Mm) so that the total coronal \lq\lq loop\rq\rq\ length is $100$~Mm, with a corresponding TR ($z=100$~Mm), chromosphere ($100 < z < 106$~Mm), photosphere ($z = 106$~Mm) and convection zone ($z > 106$~Mm) at the top boundary.
In this model, the two photospheres correspond to the negative and positive polarity regions of an active region. 
Figure \ref{fig:hydro_atmosphere} shows that the plasma $\beta=1$ at the photospheres, reduces to $\beta=10^{-3}$ in the corona, while $\beta>1$ in the convection zones.

The convection zone to chromosphere parts of this hydrostatic state are thermodynamically maintained by smoothly reducing the optically thin radiative losses to zero over a 200~K range above the chromospheric temperature of $2\times10^4$~K \citep{Klimchuk1987,Bradshaw2013}.
Meanwhile, in the TR and corona, there is no form of \lq\lq background\rq\rq\ heating imposed to balance the energy losses from thermal conduction and optically thin radiation. 
Thus, the coronal plasma initially cools and drains until sufficient nanoflare heating is established to self-consistently heat the corona.

Periodic boundary conditions are enforced at the side boundaries in $X$ and $Y$.
At the bottom and top boundaries in $Z$, all velocity components and normal gradients of $B_x$ and $B_y$ are set to zero and $B_z$ is determined to ensure $\nabla \cdot {\bf B}=0$, while density and pressure are calculated to enforce hydrostatic equilibrium. 
These bottom and top boundary conditions are applied in the convection zones at $z = -15$ and $z=115$~Mm, respectively. 
However, as described in detail in the next section, we also impose a prescribed velocity driver at the two photospheric boundaries to twist and tangle the magnetic field lines.
\begin{figure}
    \hspace{2mm}
    \includegraphics[width=0.4\textwidth]{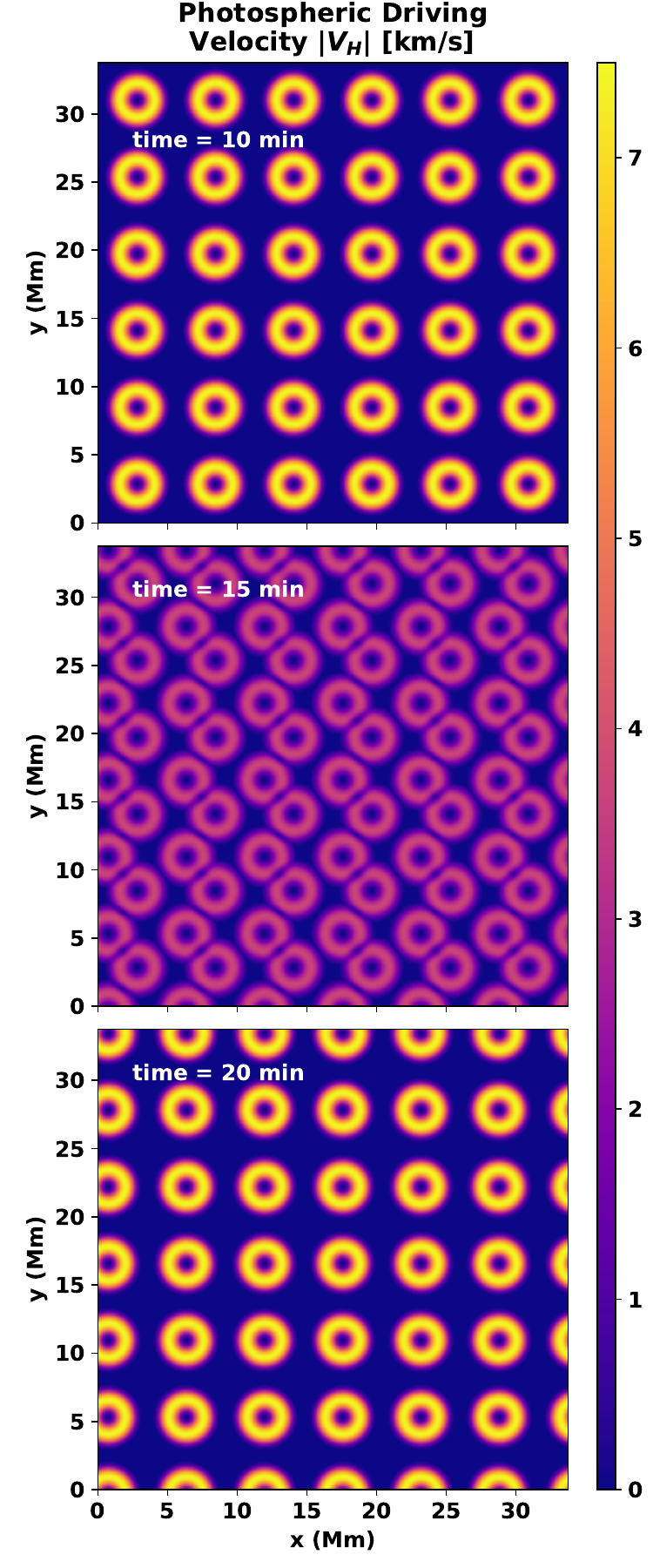}
    \vspace{-3mm}
    \caption{Magnitude of the horizontal velocity on the bottom photospheric boundary at three different times during a driving period.
    The top and bottom panels show times when each of the two $6 \times 6$ arrays reach their maximum value, while the other is at minimum. 
    The middle panel shows an intermediate time when the two  arrays temporally overlap.
    }
  \label{fig:v_driver_cells} 
\end{figure}
\begin{figure}
    \hspace{-4mm}
    \includegraphics[width=0.5\textwidth]{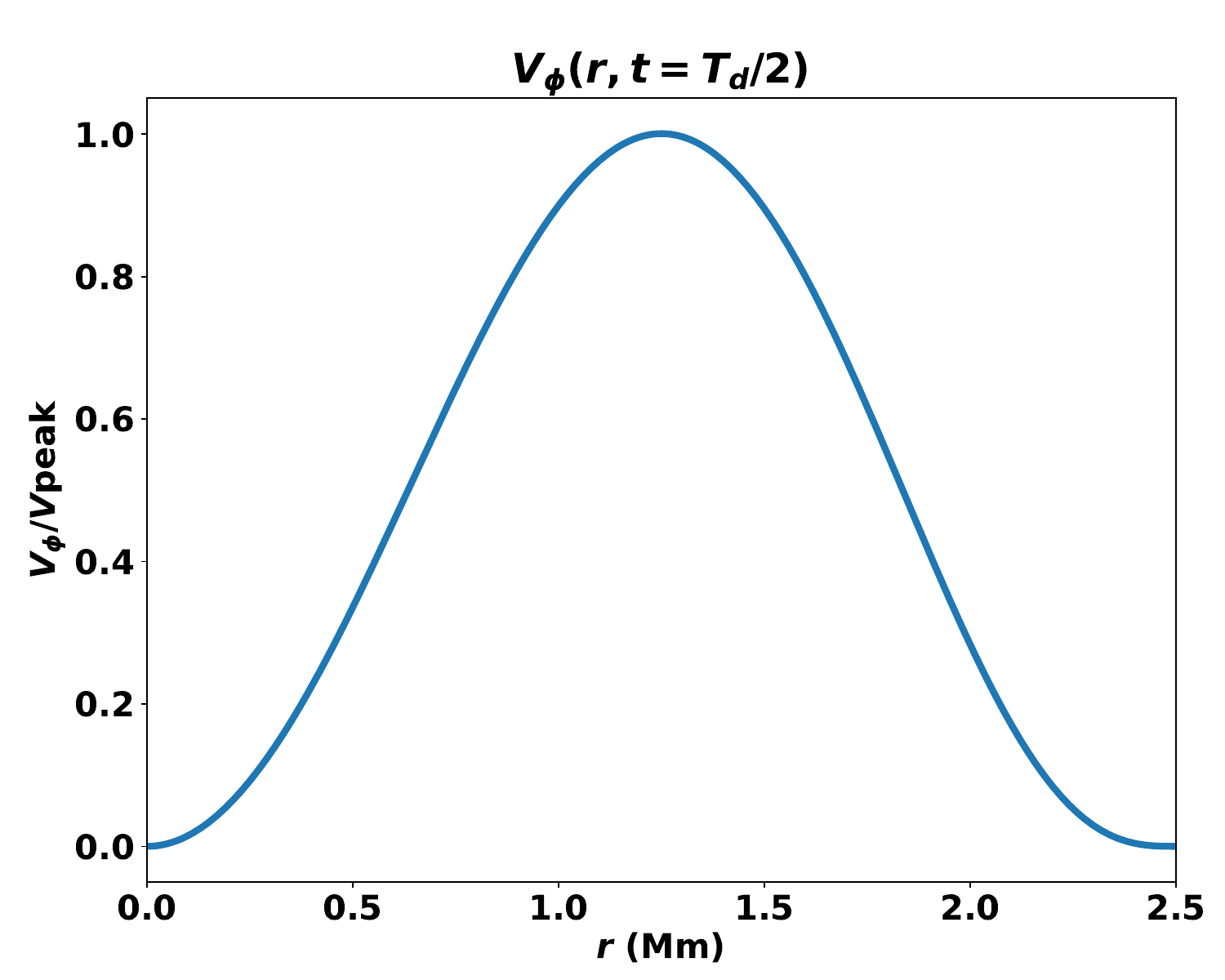}
    \vspace{-6mm}
    \caption{Radial profile of a single rotational cell at a time of maximum velocity normalized by $v_{\textrm{peak}}=7.5$~km/s.}
  \label{fig:v_phi} 
\end{figure}
\subsection{Photospheric Driving Velocity}

We drive the magnetic field in each photosphere using a velocity profile that consists of two $6 \times 6$ arrays of rotational cells.
Each array fades in and out over a $20$~min period, before being translated randomly at the end of each period. 
The two arrays are half a period out of phase so that the average magnitude of the horizontal velocity in the photosphere remains constant in time, while the maximum value peaks every $10$~min.
A net magnetic helicity is injected by imposing cells with the opposite sense of rotation at the bottom and top photospheric boundaries. 

The centers of the 36 rotational cells in each of the $6 \times 6$ arrays are uniformly arranged in $X$ and $Y$ as shown in the top and bottom panels of Figure \ref{fig:v_driver_cells}.
Each cell has a radius of $r_0=2.5$~Mm, and the separation between each cell is $s=0.625$~Mm so that $12 r_0 + 6s = 33.75$~Mm, which is the width of the box.

To randomly translate the driving pattern at the end of each driving period, we displace the center point of the $6 \times 6$ array in $X$ and $Y$, independently, by a random distance between 0 and $2r_0 + s$. 
Periodic boundary conditions are then enforced on the resulting translation to ensure a continuous driving pattern.
The radius of each of the 36 rotational cells and the separation between them remains unchanged after each translation.
However, the middle panel of Figure \ref{fig:v_driver_cells} demonstrates that the percentage of the photospheric area which is subjected to driving is increased when the two $6 \times 6$ arrays temporally overlap, with a spatial departure between the center points of their arrays. 

Within each of the $6 \times 6$ arrays, each rotational cell has a radial profile of azimuthal velocity of the form:
\begin{align}
    v_{\phi}(r,t) = v_0 f(r) D(t),
    \label{Eqn:v_phi}
\end{align} 
where $r=\sqrt{(x-x_c)^2+(y-y_c^2})$ for a rotational cell centered at ($x_c$, $y_c$), $v_0$ is a parameter that governs the rotation speed, $f(r)$ is the radial profile of the rotation and $D(t)$ is the temporal profile associated with the phasing of the two $6 \times 6$ arrays.

We specify the radial profile of each rotational cell to be incompressible with the form:
\begin{align}
    f(r) =  
    \frac{r^2}{r_0^2}
    \left(
    1 - 
    \frac{r^2}{r_0^2}
    \right)^{\!\! 3},
    \label{Eqn:f_r}
\end{align} 
which is shown in Figure \ref{fig:v_phi}. 
Consistent with the details described above, $r_0=2.5$~Mm is the radius of the cell. 
Therefore, our numerical resolution supports 38 grid points across the $5$~Mm diameter of each cell.
Furthermore, the ratio of coronal \lq\lq loop\rq\rq\ length to driver cell diameter is of order 20, which is significantly closer to typical solar values than the ratios used previously by \cite{Knizhnik2018} and \cite{Klimchuk2023a}.

The temporal profile of the two $6 \times 6$ arrays is specified to have the form:
\begin{equation}
    D(t) = 
   \begin{cases}
    \frac{1}{2}
    \left[
    1 - 
    \cos 
    \left( 
    \frac{2\pi t}{T_d}
    \right)
    \right], 
    & \textrm{if } t \geq 0, 
    \\
    \vspace*{-3mm}
    \\
    \frac{1}{2}
    \left[
    1 - 
    \cos 
    \left( 
    \frac{2\pi (t-T_d/2)}{T_d}
    \right)
    \right],  
      & \textrm{if } t \geq T_d/2,
    \end{cases}
    \label{Eqn:D_t}
  \end{equation}
where $T_d=20$~min is the period of the driving and the phasing of the arrays is such that the first array starts at $t=0$~min, while the second array starts halfway through the first period at $t=T_d/2$.
\begin{figure}
    \hspace{-4mm} 
    \includegraphics[width=0.5\textwidth]{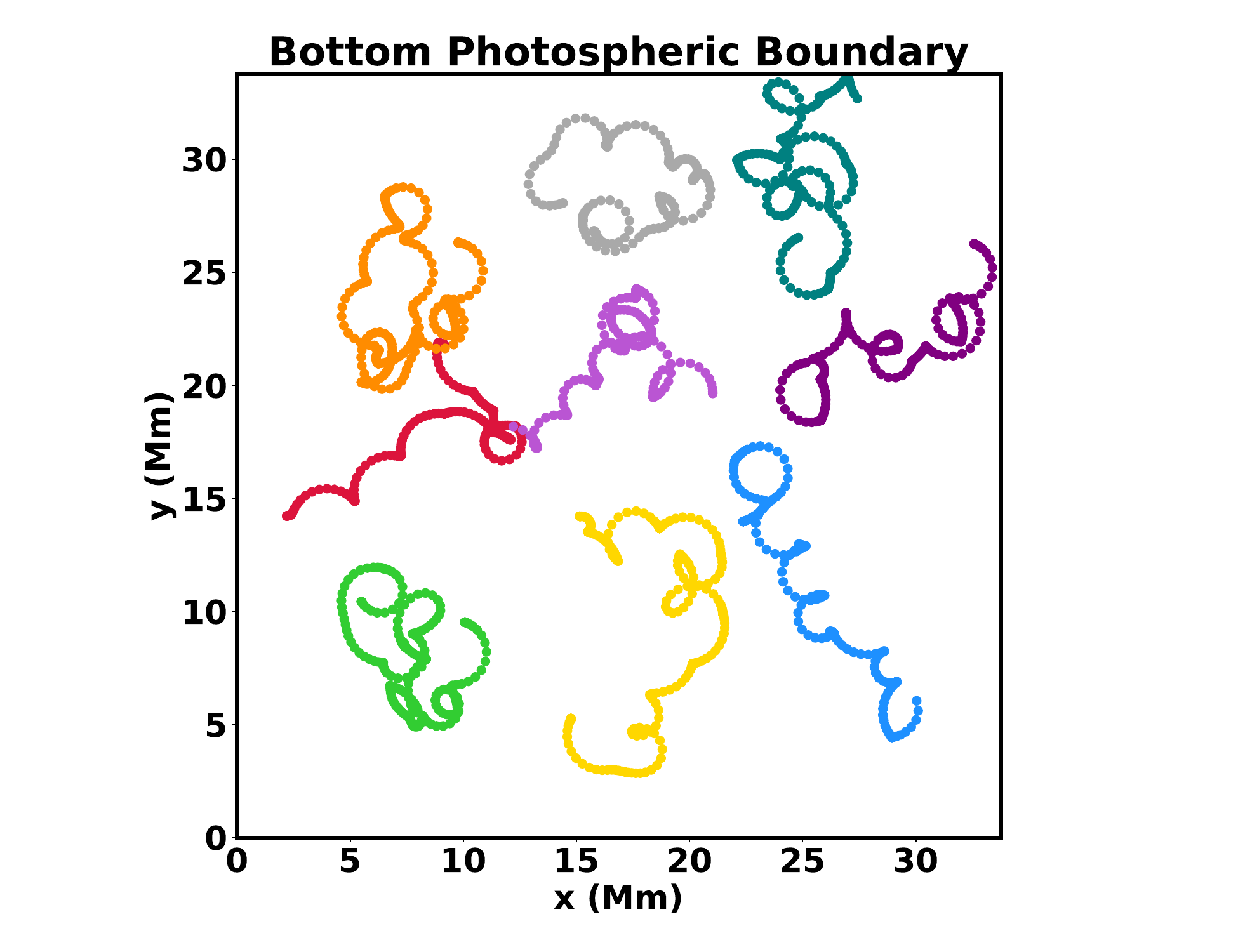}
    \vspace{-6mm}
    \caption{Trajectories of 9 randomly selected footpoints in the bottom photospheric boundary over 15 periods of driving from $t=0-300$~min.}
  \label{fig:fp_trajectories} 
\end{figure}

Each rotational cell has a maximum horizontal velocity of $v_{\textrm{peak}}=7.5$~km/s, which corresponds to $v_0=72$~km/s in Eq. \eqref{Eqn:v_phi}, while the average magnitude of the horizontal velocity in the photosphere is $\approx 2$~km/s. 
This magnitude of the driving speed is slightly faster than observed values \citep[see e.g.,][]{Rieutord2010}, but remains within a realistic range, and was chosen to ensure that all footpoint motions are both sub-sonic and sub-Alfvénic.
We also note that the chromosphere-to-chromosphere Alfvén travel time is approximately 20~s, compared with the driver period of 20~min (1200~s) ensuring that the coronal evolution in the simulation is effectively quasi-static except during reconnection events.

Figure \ref{fig:fp_trajectories} shows the trajectories of nine randomly selected footpoints in the bottom photospheric driving plane. The dots are spaced one minute apart and show the first 300 minutes of driving. 
Representative of photospheric convection, the resulting driving pattern resembles a random walk with curved steps, showing both translational and rotational components.

This photospheric driving is imposed inside the computational domain at the two photospheric boundaries positioned at $z = -6$~Mm and $z = 106$~Mm, respectively, by using  driving layers that extend downwards from these boundaries into the convection zones, as shown by the shaded regions in Figure \ref{fig:hydro_atmosphere}.
In particular, to minimize any numerical diffusion of the imposed pattern, the horizontal velocity is driven uniformly at 5 stencil heights used by the finite-difference scheme (15 grid points), while the vertical velocity is set to zero.
Only the velocities are prescribed throughout these driving layers; the remaining variables are solved according to the MHD Eqs. \eqref{Eqn:mhd_continuity}--\eqref{Eqn:mhd_gas_law}.
Since the imposed flow is analytically incompressible, the vertical component of the magnetic field remains uniform to a high degree in the driving layers, while the horizontal field evolves in response to the footpoint motions.
This approach allows us to retain consistency between the imposed velocity and subsequent evolution of the magnetic field at the photosphere, avoiding spurious boundary condition effects that plague some other simulations (see e.g., Daldorff et. al, in preparation). 
Furthermore, diffusion is applied below the driving layers, in the convection zones, to prevent small scale structures constraining the Courant time step in those particular parts of the domain.
The outcome is an efficiently controlled photospheric driver for the domain of interest, which is comprised of the region between the two photospheric boundaries ($-6 < z < 106$~Mm). 

\begin{figure}
    \includegraphics[width=0.23\textwidth,viewport=400 10 1100 1350,clip=]{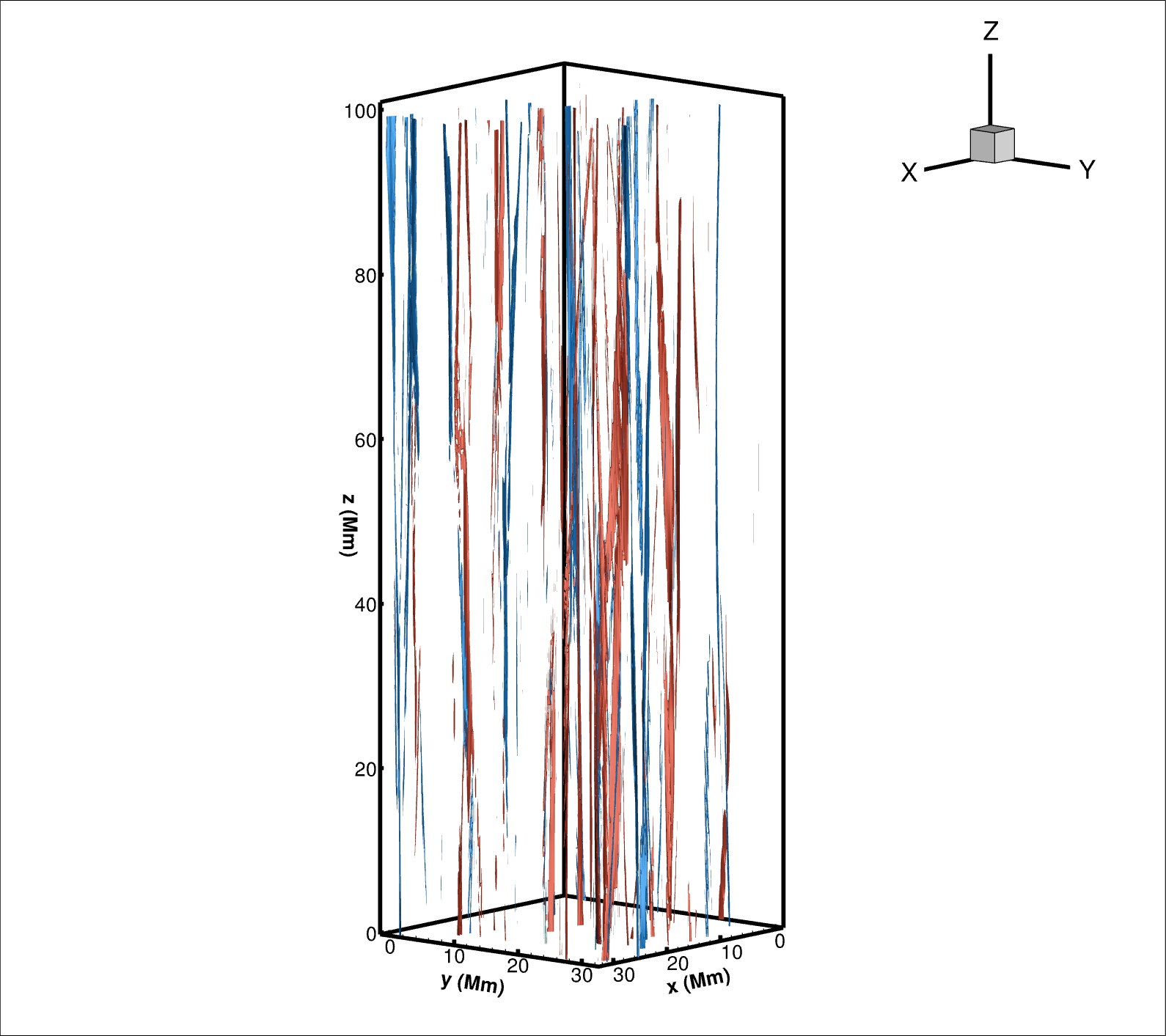}
    %
    %
    %\hspace{5mm}
    %
    %
    \includegraphics[width=0.23\textwidth,viewport=400 10 1100 1350,clip=]{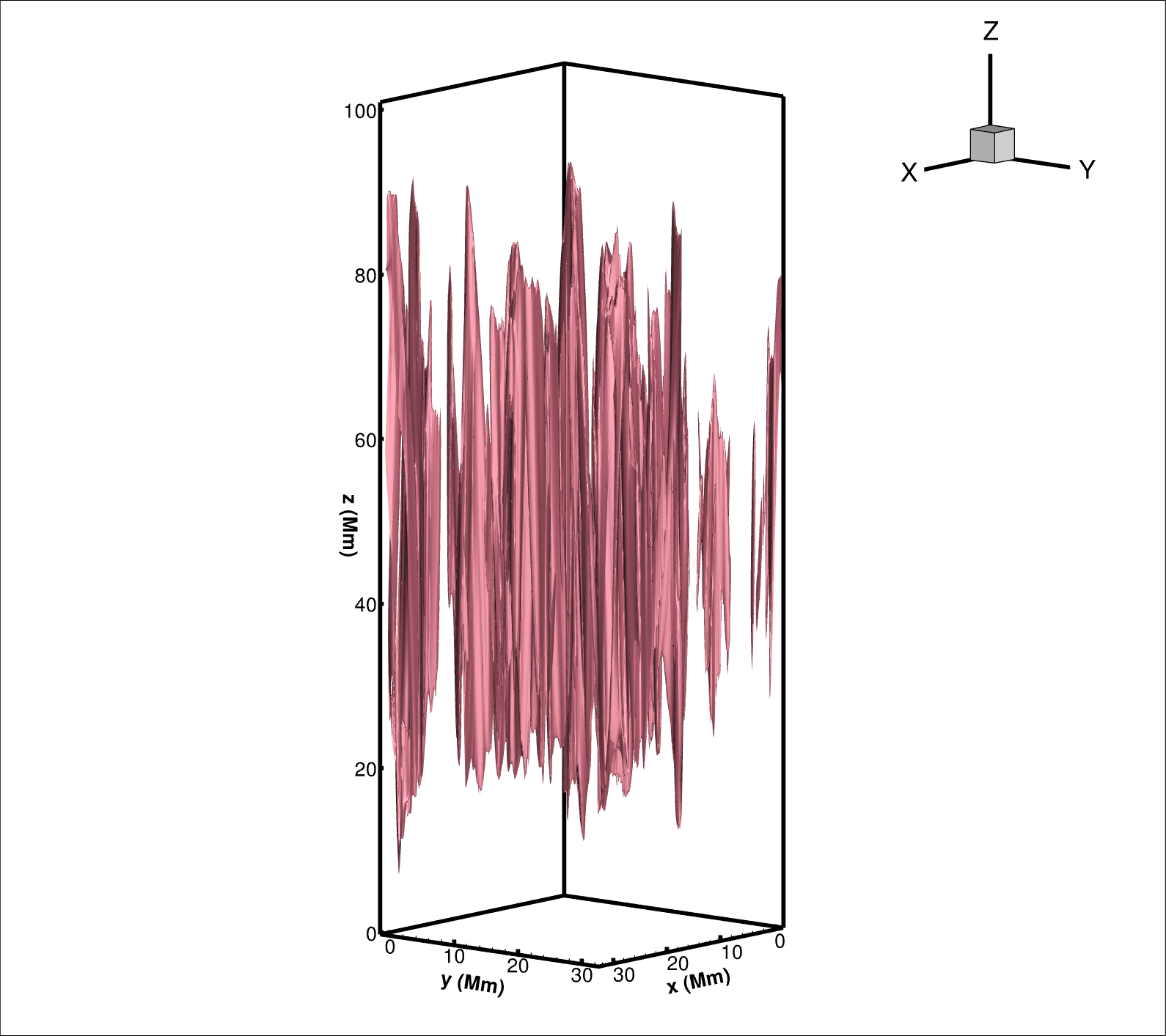}
    \vspace{-1mm}
    \caption{Left: iso-surfaces of $J_z$ current density at $\pm 2 \times 10^{-3}$~Am$^{-2}$, showing the many thin elongated current sheets that have formed at $t=300$~min.
    Blue/red corresponds to negative/positive $J_z$.
    Right: iso-surfaces of temperature at $t=300$~min, showing coronal plasma at $2$~MK.
    }
  \label{fig:current_sheets} 
\end{figure}

The imposed photospheric motions twist and tangle the magnetic field, gradually building up magnetic energy in the corona.
The left panel of Figure \ref{fig:current_sheets} shows the many thin current sheets that form at the boundaries between quasi-homogeneous magnetic flux strands. 
These current sheets subsequently facilitate reconnection and the rapid release of magnetic energy as nanoflares that impulsively heat the coronal plasma, as shown in the right panel of Figure \ref{fig:current_sheets}. 
Eventually the system reaches a statistical steady state, where the injected Poynting flux balances the energy released during the impulsive heating events.
Then finally after 15 periods of driving, at $t=300$~min, all symmetries in the system have been destroyed and the untwisted initial state forgotten.
Thus, in the remainder of this paper, we present the results of the simulation starting at t = 300 min and analyze the evolution over 15 further periods of driving up to $t = 600$~min.
  \begin{figure*}
    {\textbf{(a)}}
    \\[-0.4mm]
    \includegraphics[width=0.97\textwidth]{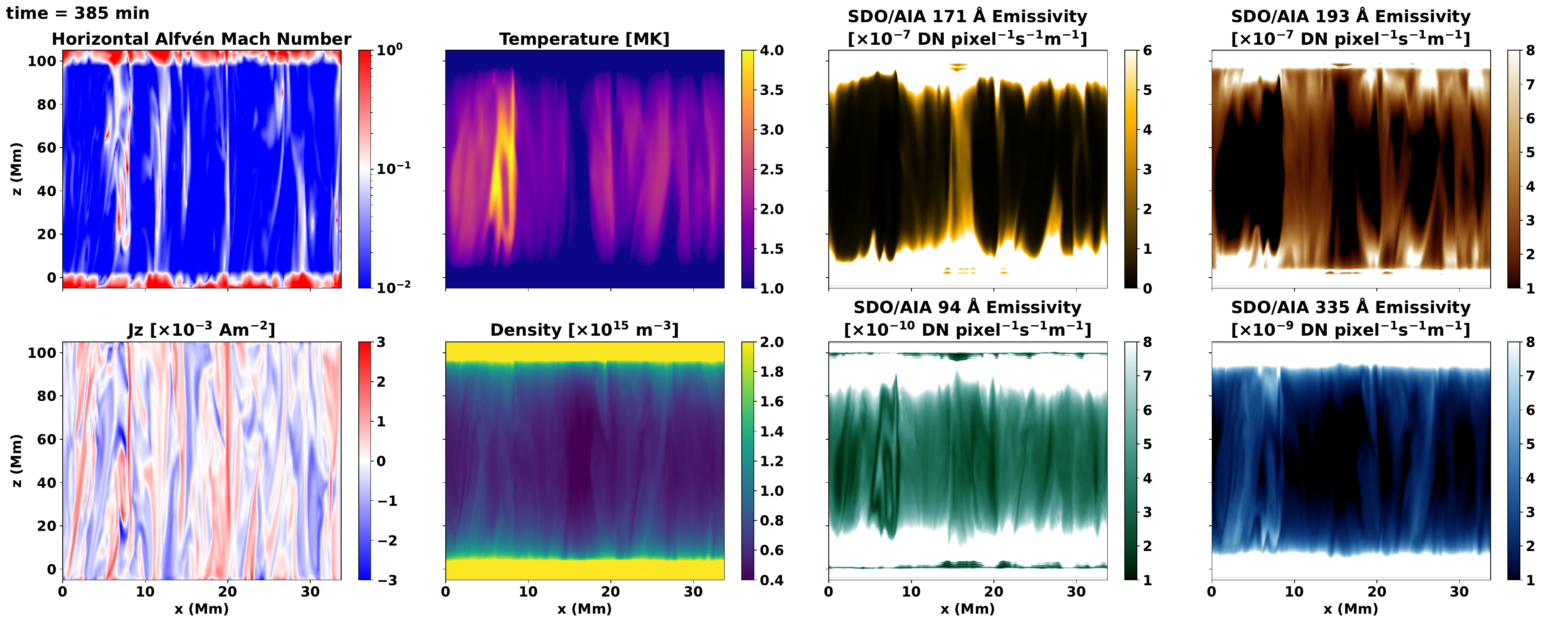}
    \\
    {\textbf{(b)}}
    \\[-0.4mm]
    \includegraphics[width=0.97\textwidth]{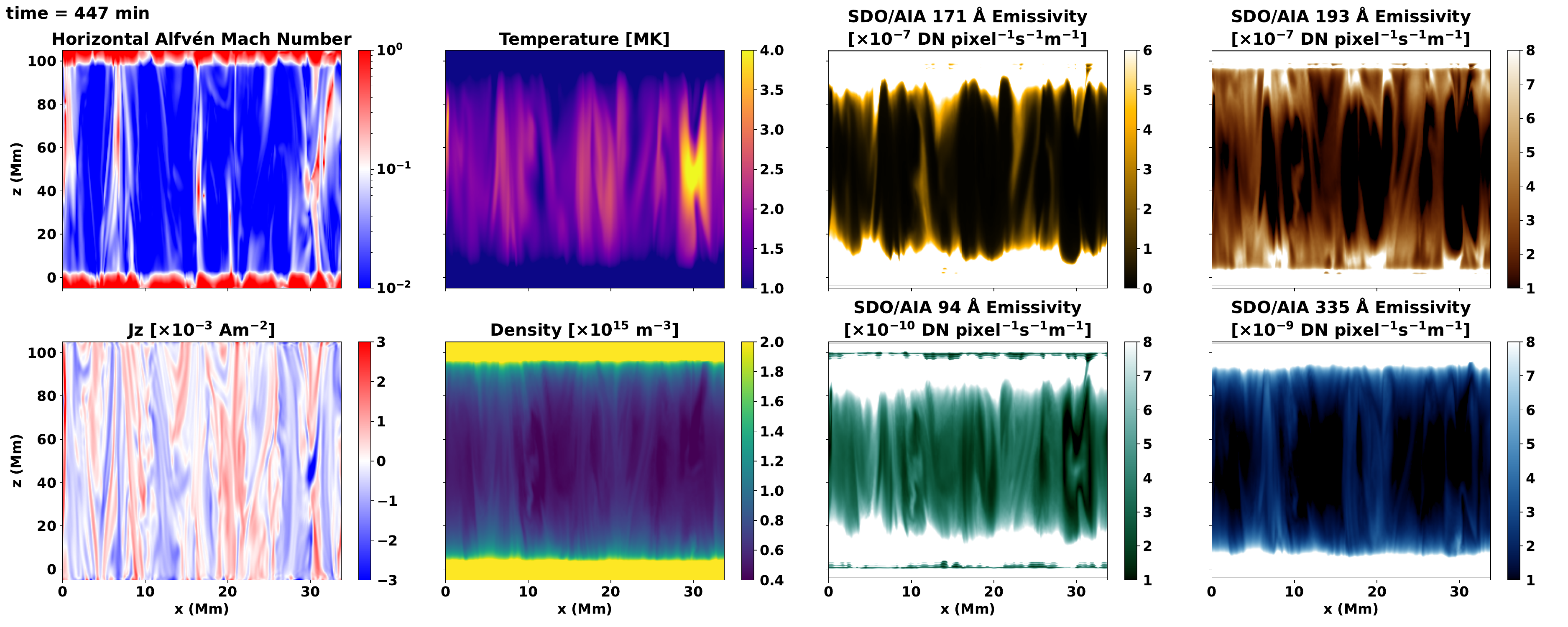}
    \\
    {\textbf{(c)}}
    \\[-0.4mm]
    \includegraphics[width=0.97\textwidth]{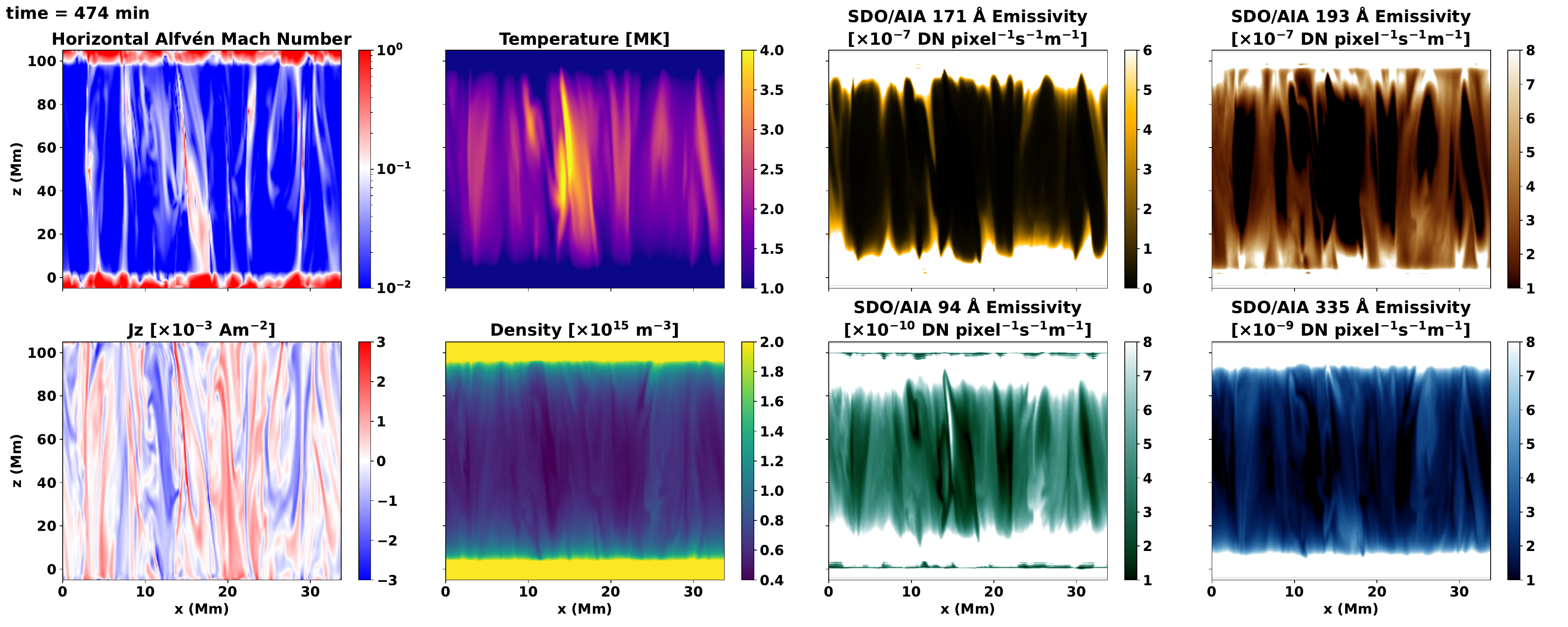}
    \caption{Time ordered snapshots showing the generation of nanoflares and response of the plasma in the $Y$-midplane at three different times.
    The contours are drawn according to the scales shown in the colour tables.
    For each snapshot the left four panels show the horizontal Alfv\'en Mach number, temperature, $J_z$ current density and electron number density in the $Y$-midplane, and the right four panels show the emissivities that would be detected in the 171~{\AA}, 193~{\AA}, 94~{\AA} and 335~{\AA} channels of AIA.
    A movie of the full time evolution from $t=300-600$~min can be viewed online.
\\
    }
    \label{fig:emissivity_midplane_xz}
  \end{figure*}
\begin{figure}
    \hspace{-4mm}
    \includegraphics[width=0.5\textwidth]{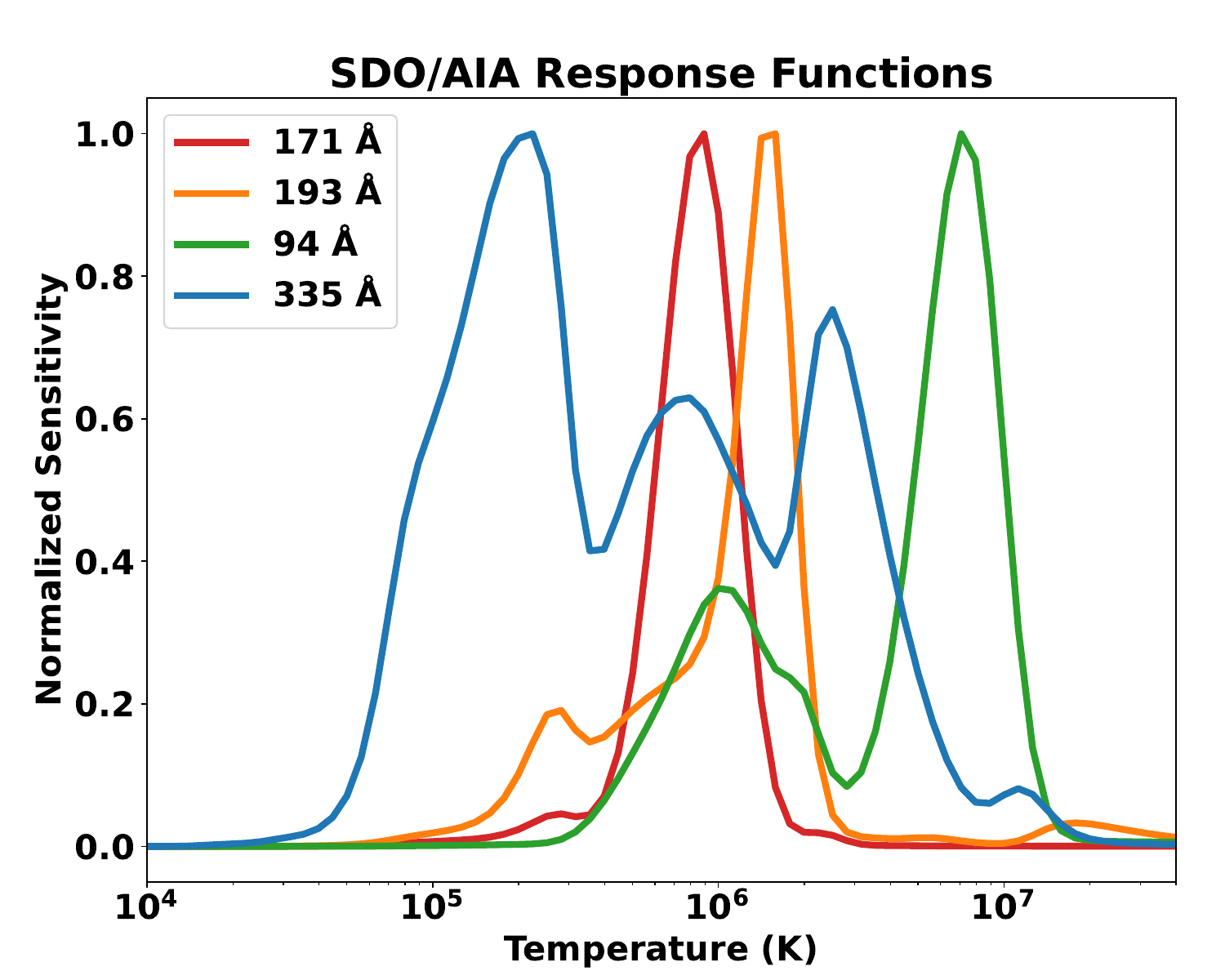}
    \vspace{-6mm}
    \caption{SDO/AIA temperature response functions for the four channels used throughout this paper: 171~{\AA} (red), 193~{\AA} (orange), 94~{\AA} (green) and 335~{\AA} (blue).
    }
  \label{fig:aia} 
\end{figure}
%
%
%
%
%%%%%%%%%%%%%%%%%%%%%%%%%%%%%%%%%%%%%%%%%%%%%%%%%%%%%%%%%%%%%%%%%%%%%%%%%%%%%%%%%%%%%%%%%%%%%%%
%
%
\section{Results} \label{sec:results}

\subsection{Nanoflares and Response of the Plasma}

Figure \ref{fig:emissivity_midplane_xz} shows the generation of nanoflares and response of the plasma in the $Y$-midplane of the simulation.
This plane is vertically stratified and so the coronal plasma extends along the magnetic field between the TR, chromosphere and photosphere that are present at the top and bottom boundaries.
Snapshots at three different times are shown in the panels to demonstrate typical nanoflare events: (a) $t = 385$~min, (b) $t = 447$~min and (c) $t = 474$~min. 
For each snapshot maps of the horizontal Alfv\'en Mach number ($M_{A_H}=v_{_H}/v_{A_H}$ where 
$v_{_H}=\sqrt{v_x^2+v_y^2}$ and 
$v_{A_H}=\sqrt{B_x^2+B_y^2}/\sqrt{\mu_0\rho}$), $J_z$ current, temperature and density in the midplane are presented, together with the corresponding maps of emissivity that would be detected in the 171~{\AA}, 193~{\AA}, 94~{\AA} and 335~{\AA} channels of AIA.
In particular, the $J_z$ current and horizontal Alfv\'en Mach number plots demonstrate the generation of the nanoflares by showing the sites of reconnection and resulting locations of shock heating, respectively.
The temperature and density plots show the plasma response to these nanoflares and the emissivity maps present the predicted AIA emission for this nanoflare heating. 
A movie covering the full time evolution of the simulation, using the same visualization, can be viewed online.

We note that the emissivity maps appear artificially bright in the chromosphere, photosphere, and convection zone because they were computed under the assumption of optically thin emission. However, these dense regions are expected to optically thick, and so the actual emission would be greatly diminished. As such we confine our analysis to the much less dense transition region and corona ($0< z < 100$~Mm), where this optically thin emission is physically meaningful.

The three snapshots shown in Figure \ref{fig:emissivity_midplane_xz} and the movie online reveal many impulsive heating events that are distributed throughout the $Y$-midplane.
These nanoflares heat the coronal plasma to temperatures exceeding 4~MK. This directly heated plasma is always co-spatial with horizontal Alfv\'en Mach numbers of order 1, corresponding to regions with strong reconnection outflow jets.
For example, the nanoflare seen in panel (b) of Figure \ref{fig:emissivity_midplane_xz}, at $t = 447$~min, is generated by viscous dissipation of the outflow jets from the reconnecting current sheet that is associated with the X-shaped crossing strands, which have formed on the right-hand side as a result of the imposed photospheric motions.
The response of the plasma to this nanoflare then follows the familiar evaporation process \citep{Antiochos1978,Klimchuk2006,Klimchuk2008,Cargill2012a,Reale2014}.
In particular, the thermalized magnetic energy released by the nanoflare gets conducted downwards along the reconnected field lines into the TR and chromosphere, where it increases the gas pressure locally \citep{Johnston2019a}.
The resulting pressure gradients then drive upflows of material that increase the coronal density. 
After the time of peak density, this process is reversed and the material drains in order to power the TR radiation \citep{Bradshaw2010a,Bradshaw2010b}.

Similar characteristics are also shared by the other two nanoflare events that are presented in panels (a) and (c) of Figure~\ref{fig:emissivity_midplane_xz}. However, these heating events also demonstrate some subtle differences in the details of the reconnection that leads to their energy release.
Firstly, the nanoflare at $t = 385$~min is associated with reconnection at multiple current sheets that have formed at different heights throughout the plane, whereas the nanoflare at $t = 447$~min has one main current sheet.
Secondly, the reconnection that is responsible for the nanoflare at $t = 474$~min occurs in a current sheet that is more elongated in height than the localized current sheet that is associated with the  nanoflare at $t = 447$~min.

A range of different nanoflares can also be identified throughout the evolution seen in the movie. 
Collectively, these nanoflares and the resulting temperature and density response of the plasma determine the brightness of the predicted AIA emission. 
As shown in Figure~\ref{fig:aia}, the four AIA channels used have peak sensitivities at the following temperatures: 0.8~MK~(171~{\AA}), 1.6~MK~(193~{\AA}), 6.4~MK~(94~{\AA}) and 2.5~MK~(335~{\AA}). 
For the 171~{\AA}, 193~{\AA} and 335~{\AA} channels the predicted emission is produced predominantly by cooler evaporated plasma, which is emitting at coronal temperatures that correspond to the peak sensitives of these particular channels (0.8--2.5~MK).
However, the AIA filters are sensitive to a broad range of temperatures and notably
the 94~{\AA} channel has a secondary peak at 1~MK. 
Therefore, the emission produced in the 94~{\AA} channel is comprised of contributions from both the hot directly heated plasma (6.4~MK) and cooler evaporated plasma (1~MK). 

Contrasting properties are evident for these two parts of the predicted emission.
On one hand, the emission from the hot directly heated plasma is extremely faint because of its low density, and it is  short-lived due to highly efficient conductive cooling along the magnetic field \citep{Bradshaw2006}. 
On the other hand, the emission from the cooler evaporated plasma dominates the emissivity maps because of its higher density, and it is longer-lived due to slower radiative cooling to space \citep{Bradshaw2010a,Bradshaw2010b}. 
Thus, the simulation produces an abundance of emission at coronal temperatures between 0.8--2.5~MK, but with extremely low count rates for plasma above 6~MK.
  \begin{figure*}
    {\textbf{(a)}}
    \\[-0.4mm]
    \includegraphics[width=0.97\textwidth]{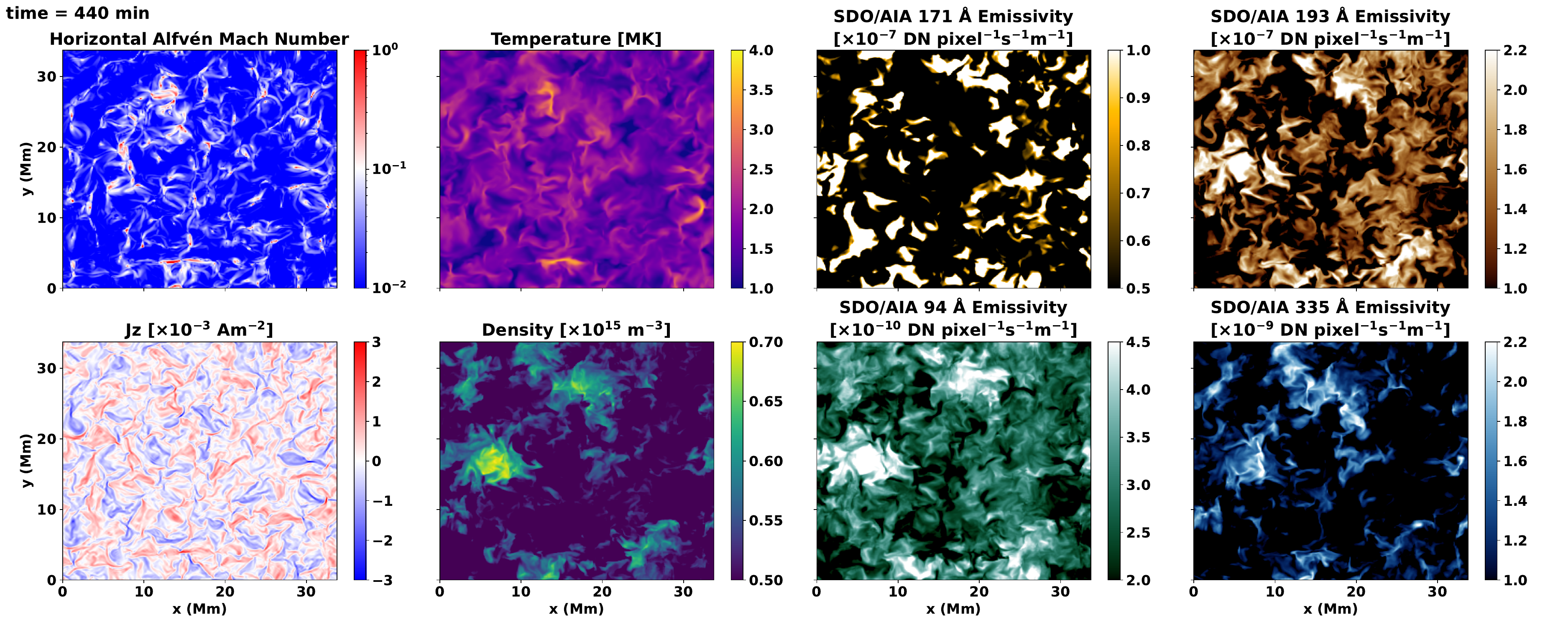}
    \\
    {\textbf{(b)}}
    \\[-0.4mm]
    \includegraphics[width=0.97\textwidth]{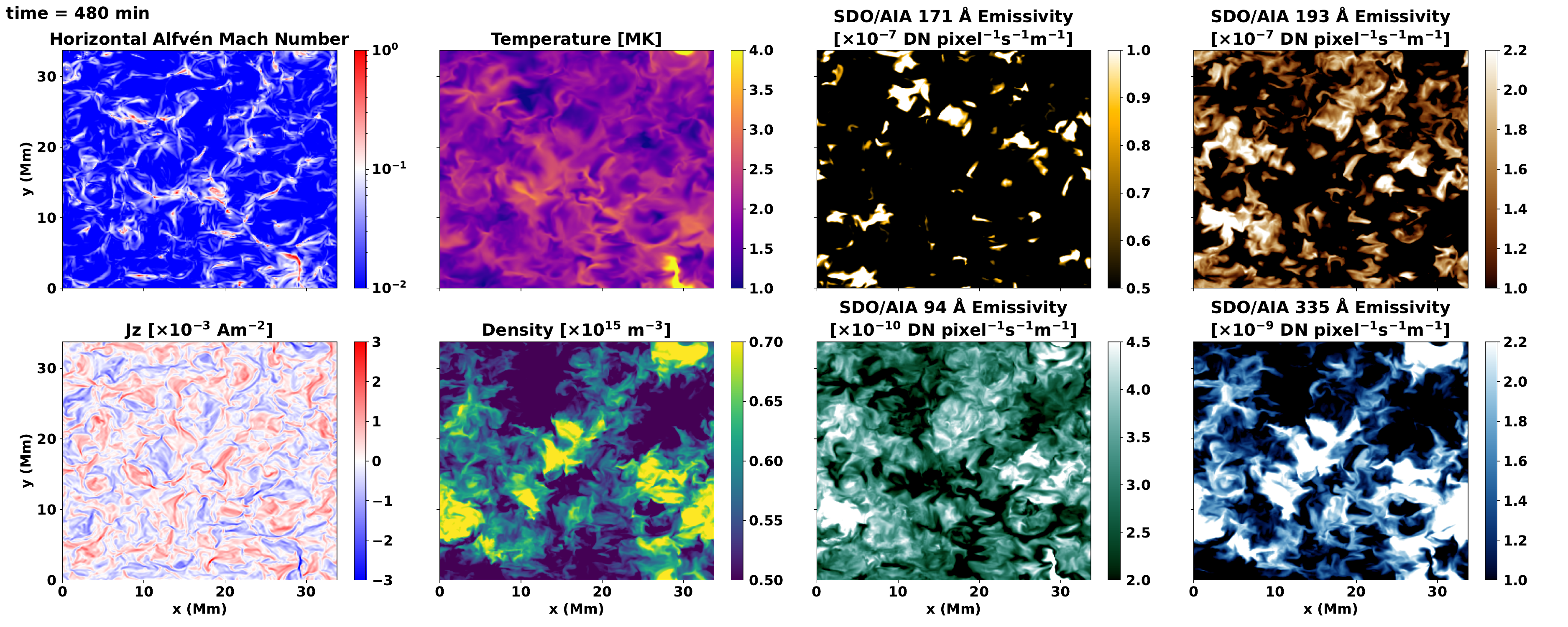}
    \\
    {\textbf{(c)}}
    \\[-0.4mm]
    \includegraphics[width=0.97\textwidth]{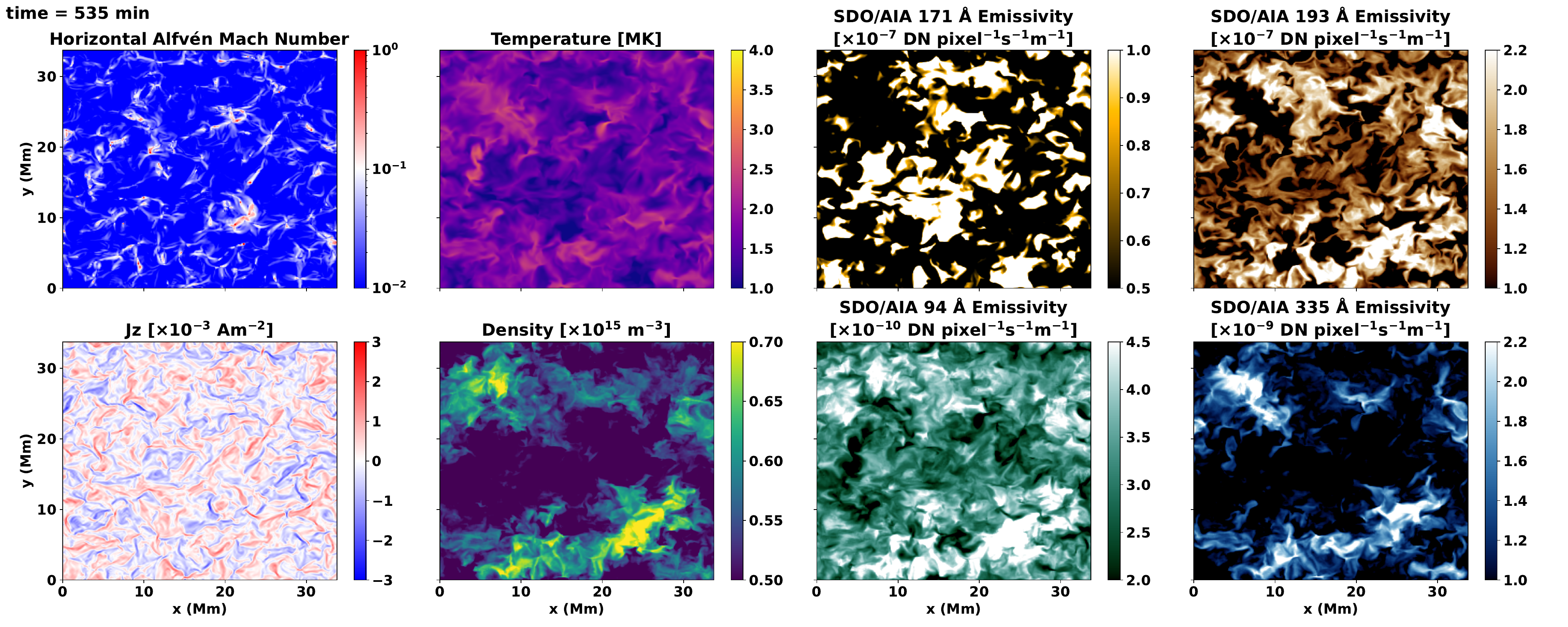}
    \caption{Nanoflares and emission produced in the $Z$-midplane.
    Notation is the same as that in Figure \ref{fig:emissivity_midplane_xz}.
    A movie of the full time evolution from $t=300-600$~min can be viewed online.
\\
    }
\label{fig:emissivity_midplane_xy_clusters}
  \end{figure*}
\begin{figure*}[t]
    \centering
    \includegraphics[width=0.7\textwidth]{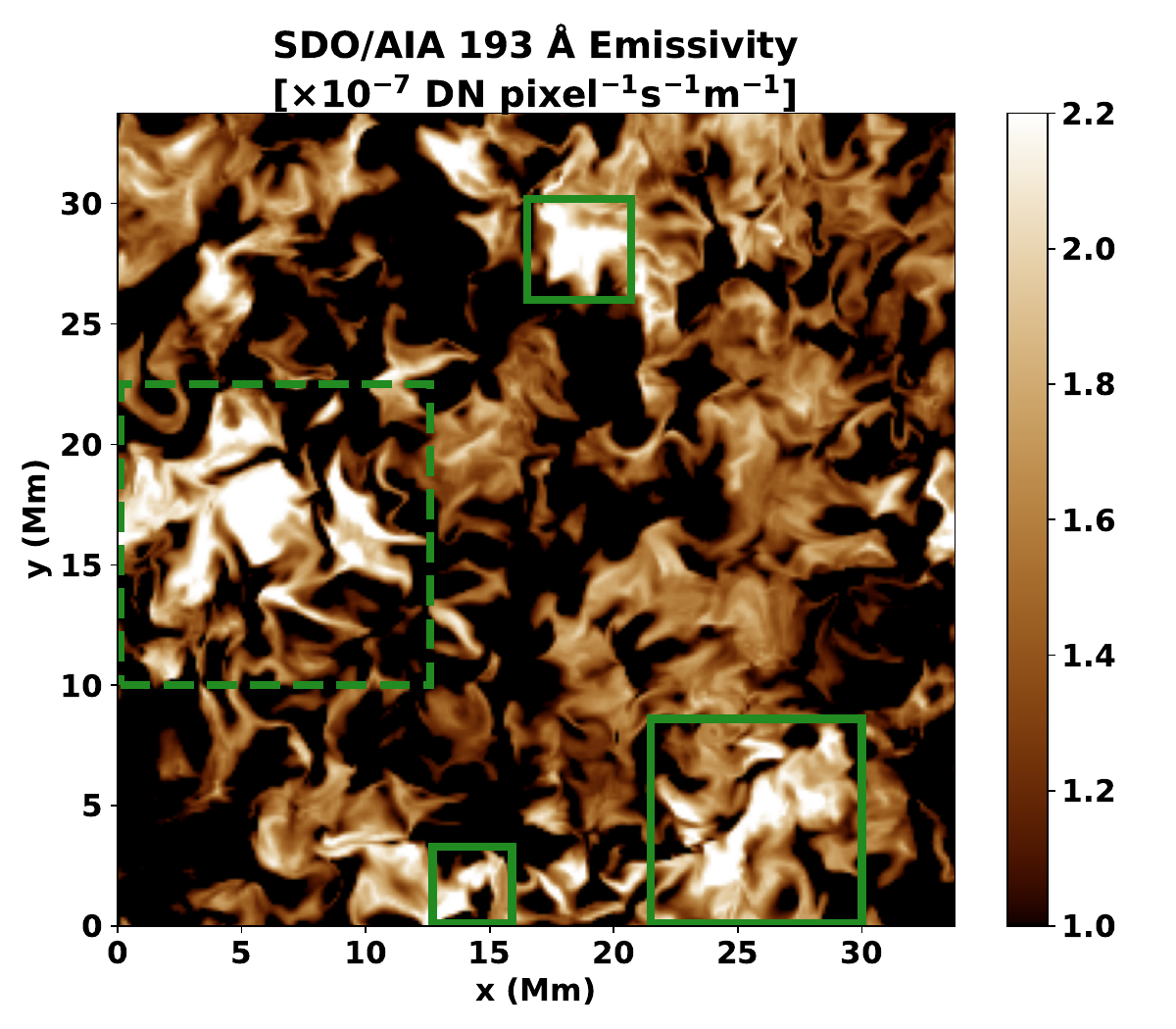}
    \vspace{-1mm}
    \caption{Emissivity in the $Z$-midplane that would be detected in the 193~{\AA} channel of AIA at $t=440$~min, showing clusters with a distribution of sizes as outlined by the green boxes.
    The dashed green box marks the largest cluster and area of the spatially averaged light curve in Figure \ref{fig:integrated_emissivity}.
    }
  \label{fig:clusters} 
\end{figure*}
\begin{figure*}
    \centering
    \includegraphics[width=0.8\textwidth]{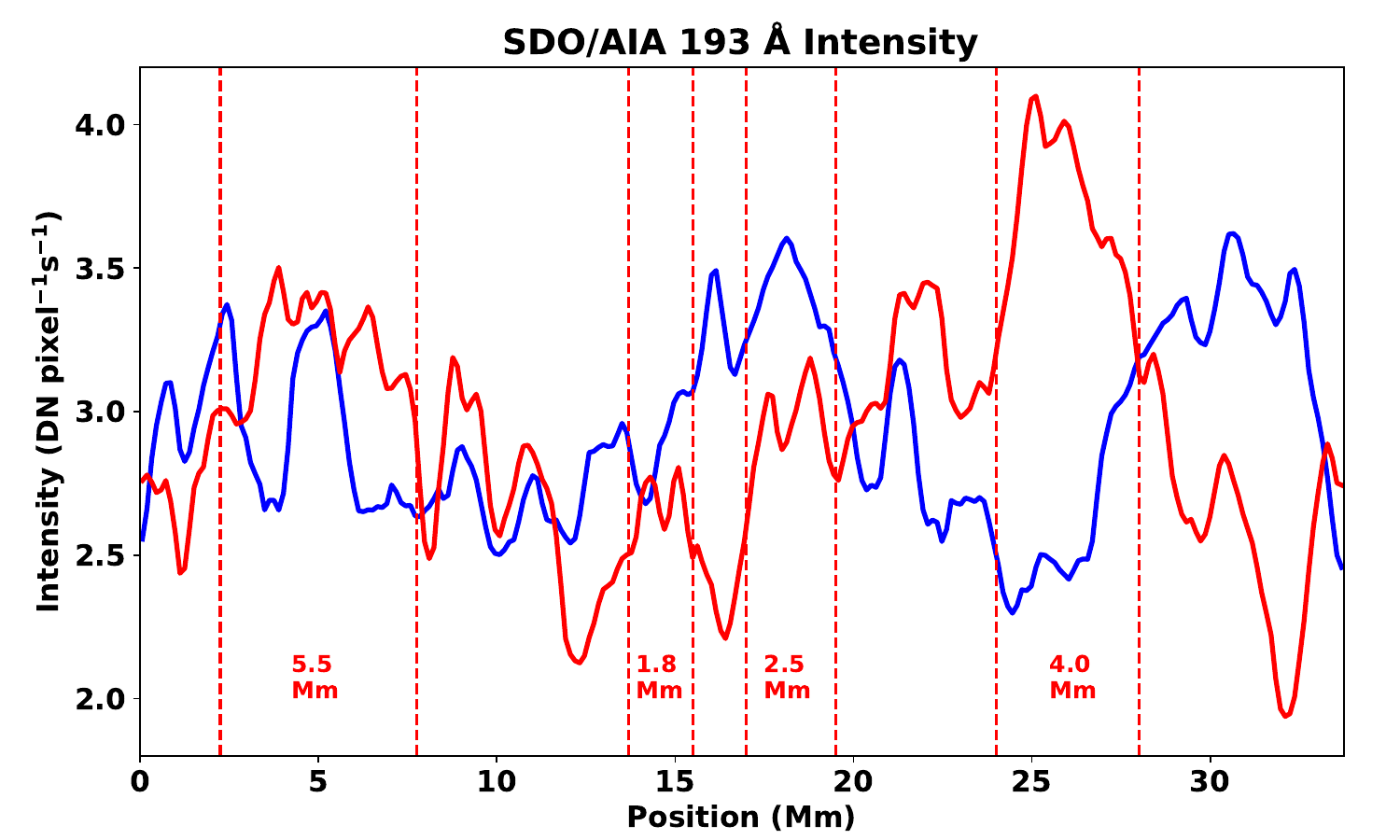}
    \vspace{-1mm}
    \caption{193~{\AA} intensity profiles (emission integrated along the LOS as a function of position) corresponding to an observation made from the top (red) and from the right (blue) of Figure \ref{fig:clusters}.
    The dashed vertical lines indicate approximate FWHMs in the red intensity profile that correspond to the clusters identified in the green boxes in Figure \ref{fig:clusters}.
    }
  \label{fig:intensity_profiles} 
\end{figure*}
\subsection{Coronal Emission}

The main question that needs to be addressed is whether the nanoflare heating in our simulation and the emission produced are supported or ruled out by coronal observations.
Figure \ref{fig:emissivity_midplane_xy_clusters} presents a detailed overview of the nanoflares and emission produced in the $Z$-midplane.
This plane shows the coronal plasma across the magnetic field at the apex of the flux tube.
Therefore, the emission produced in the $Z$-midplane can be compared against observations of coronal loop cross sections and their background emission.
The snapshots shown are taken at three different times to demonstrate the typical emission produced: (a) $t = 440$~min, (b) $t = 480$~min and (c) $t = 535$~min.

Figure \ref{fig:emissivity_midplane_xy_clusters} is accompanied by a movie online that shows the full time evolution from $t=300-600$~min.
The temporal evolution of the produced emission reveals two components that are prevalent in each of the AIA emissivity maps.
The first component is the many small, randomly scattered and irregularly shaped brightenings that give the appearance of a twinkling throughout the $Z$-midplane. 
These features are produced independently by random nanoflares.
When integrated along the line of sight, they form a diffuse background that evolves slowly even though the individual brightenings evolve rapidly.
We relate this background emission to the diffuse component of observed active regions.
The second component is the distinct clusters of brightenings that stand out above the diffuse background.
These localized clusters are roughly circular in shape and occur with a distribution of sizes, as highlighted in Figure \ref{fig:clusters}, where we identify and mark four different clusters in the 193~{\AA} channel with green boxes.
The clusters are produced by nanoflare storms that have an avalanche nature \citep{Hood2016,Reid2018,Cozzo2023}, where one event triggers subsequent events.
Thus, the lifetime of the clusters is longer than the individual features that comprise them.
We associate these clusters of enhanced emission with coronal loops.

This explanation for the coronal loops produced in our simulation is completely consistent with the picture proposed by \cite{Klimchuk2009,Klimchuk2015} to reconcile several observations that are otherwise difficult to understand. 
We explore this agreement in more detail in the next section by quantitatively comparing the results from our simulation against observational characteristics of coronal loops.
\begin{figure*}
    \centering
    \includegraphics[width=0.8\textwidth]{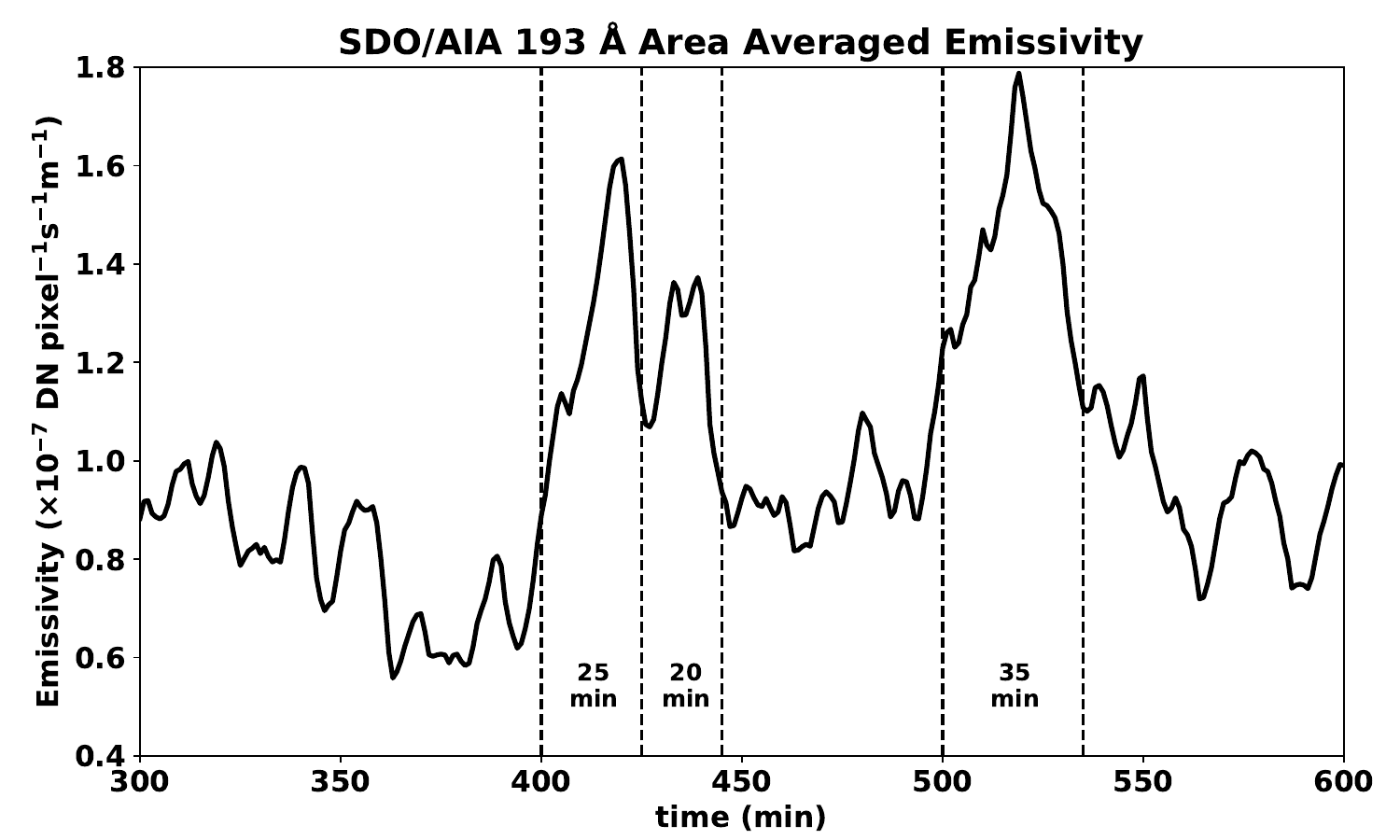}
    \vspace{-1mm}
    \caption{193~{\AA} light curve for the emissivity averaged over the dashed green box in Figure \ref{fig:clusters}.
    The dashed vertical lines indicate approximate lifetimes of three different clusters that form during the simulation.
    }
  \label{fig:integrated_emissivity} 
\end{figure*}
\subsection{Loop Widths, Lifetimes and Cross Sections}

Figure \ref{fig:intensity_profiles} shows 193~{\AA} intensity profiles, for line of sight (LOS) integrations made through the apex of the flux tube in the $Z$-midplane, at $t=440$~min.
The red (blue) curve corresponds to a hypothetical observation made from the top (right) of Figure \ref{fig:clusters} by integrating the emissivity along the LOS in $Y$ ($X$) at each position in $X$ ($Y$).
The clusters identified in Figure \ref{fig:clusters} manifest as enhancements above the background emission in the intensity profiles. 
We equate the full width at half maximum (FWHM) of these enhancements with the widths of coronal loops.
In particular, the FWHMs in the intensity profiles that are associated with the clusters range roughly between $1.8-5.5$~Mm, whilst smaller sub-elemental magnetic flux strands are also evident.
Thus, the loop widths in our simulation show reasonable agreement with the 1~Mm FWHMs that are typically reported in coronal loop observations \citep{Klimchuk2015,Williams2021,Mandal2024}.

However, we note that the rotational and translational motions in our simulation produce a random walk step size of order 5~Mm, while observations suggest that a step size closer to 1~Mm is more appropriate \citep{deWijn2008}.
Therefore, scaling the horizontal dimensions of our simulation to better match the scale of the observed photospheric driving would also further improve the agreement between the FWHMs of the clusters and observed loop widths.

Figure \ref{fig:integrated_emissivity} shows the time evolution of the 193~{\AA} emissivity averaged over the area marked by the dashed green box in Figure \ref{fig:clusters}.
There are three main rise and fall phases in the light curve.
These are associated with the lifetimes of three clusters that form at different times during the simulation.
We relate the lifetimes of these clusters to the lifetimes of coronal loops.
In particular, the first cluster forms around $t=400$~min and has a lifetime of approximately $25$~min. 
This is followed immediately by the second cluster, which is the cluster outlined by the dashed green box in Figure \ref{fig:clusters} at $t=440$~min.
The second cluster has a lifetime of about $20$~min.
Then, later in the simulation, the third cluster forms around $t=500$~min and persists for roughly $35$~min.
In comparison, the lifetime of 193~{\AA} loops in observations is typically around $30$~min, with a range between $10-300$~min frequently reported \citep{Winebarger2003,Ugarte2006}.
Thus, the loop lifetimes in our simulation show good agreement with those of observed loops.

The clusters shown in Figure \ref{fig:clusters} are all roughly circular in shape, in the sense that their aspect ratios are of order unity. 
But we note that the emissivity is not uniform within the cross sections. 
Rather the emissivity is concentrated in sub-resolution patches, which themselves can have large aspect ratios. 
However, the distribution of sub-resolution patches is approximately uniform within the cross sections to produce the circular clusters.
These circular shapes of enhanced emission are consistent with recent observational studies that have reported circular cross sections for most real loops \citep{Klimchuk2020,Williams2021,McCarthy2021,Mandal2024,Uritsky2024,Ram2025}.

Our simulation demonstrates that the circular cross sections of the clusters are a natural consequence of avalanche spread.
In particular, the magnetic flux strands are randomly twisted and tangled in such a way that the separation between any two diverging strands does not have a preferred direction.
Therefore, an avalanche propagates out radially across the magnetic field during a nanoflare storm.
This radial spread from the initial reconnection site produces the circular clusters of enhanced emission, thus providing a physical justification for the circular cross sections of observed coronal loops.

The distribution of cluster sizes is then explained by the distribution of avalanches and the cross field scale of their cascade.
Large (small) clusters are produced by avalanches with large (small) cross field cascades.
This is consistent with a continuous spectrum of reconnection events that ranges from nanoflare storms with large cross field cascades down to random nanoflares that do not trigger any subsequent events across the magnetic field.
Therefore, our simulation supports a unifying picture that explains both the diffuse component of observed active regions and the distinct coronal loops \citep{Klimchuk2009,Klimchuk2015}.

The above discussion specifically focuses on emission in the 193~{\AA} channel, but the emission in the 171~{\AA}, 94~{\AA} and 335~{\AA} channels is qualitatively similar.
Each channel has both random brightenings and clusters of brightenings.
Furthermore, many of the clusters have a multi-thermal distribution across their cross sections, producing emission in multiple channels.
This is consistent with the multi-thermal emission measure distributions that are typically reported for observed  loops \citep{Warren2008,Ugarte2009,Warren2012}.
We analyze the formation of such multi-thermal clusters and their evolution in each of the channels in the next section.
\begin{figure*}
    \vspace{-10mm}
    \centering
    \includegraphics[width=0.78\textwidth]{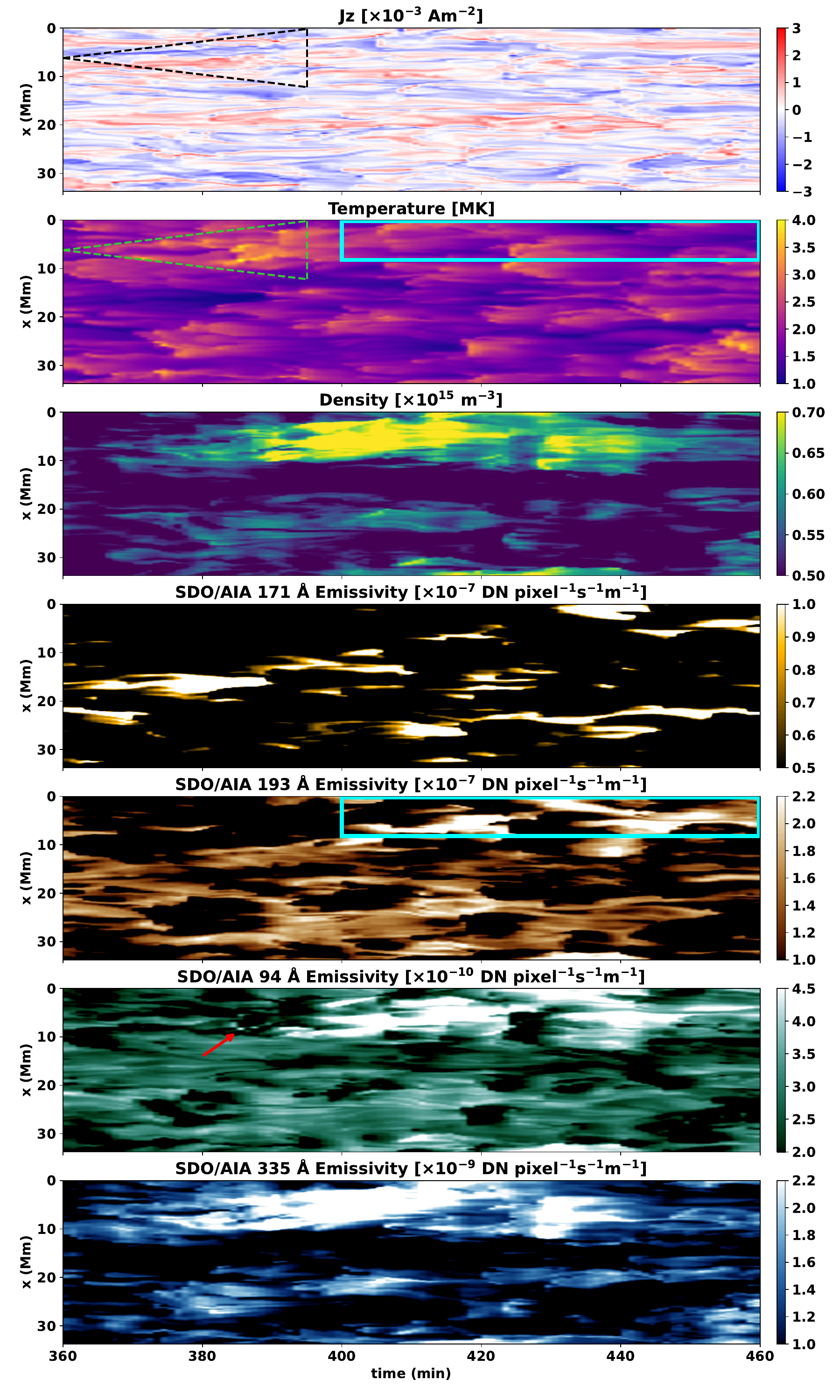}
    \vspace{-6mm}
    \caption{Time-distance plots showing a nanoflare storm and subsequent formation of multi-thermal clusters in the $Z$-midplane.
    The upper three panels show the $J_z$ current density, temperature and electron number density at $y=16$~Mm, and the lower four panels show the emissivities that would be detected in the 171~{\AA}, 193~{\AA}, 94~{\AA} and 335~{\AA} channels of AIA.
    The nanoflare storm is indicated by the dashed triangles on the $J_z$ current density and temperature panels, while the cyan boxes on the temperature and 193~{\AA} panels mark the multi-thermal, multi-stranded post-storm evolution that is shown close-up in Figure \ref{fig:193_shifts}.
    The red arrow indicates the faint signatures of the hot directly heated plasma in the 94~{\AA} channel.
    }
  \label{fig:storm} 
\end{figure*}
\subsection{Nanoflare Storms and Multi-Thermal, Multi-Stranded Emission}
\begin{figure*}
    \centering
    \includegraphics[width=0.8\textwidth]{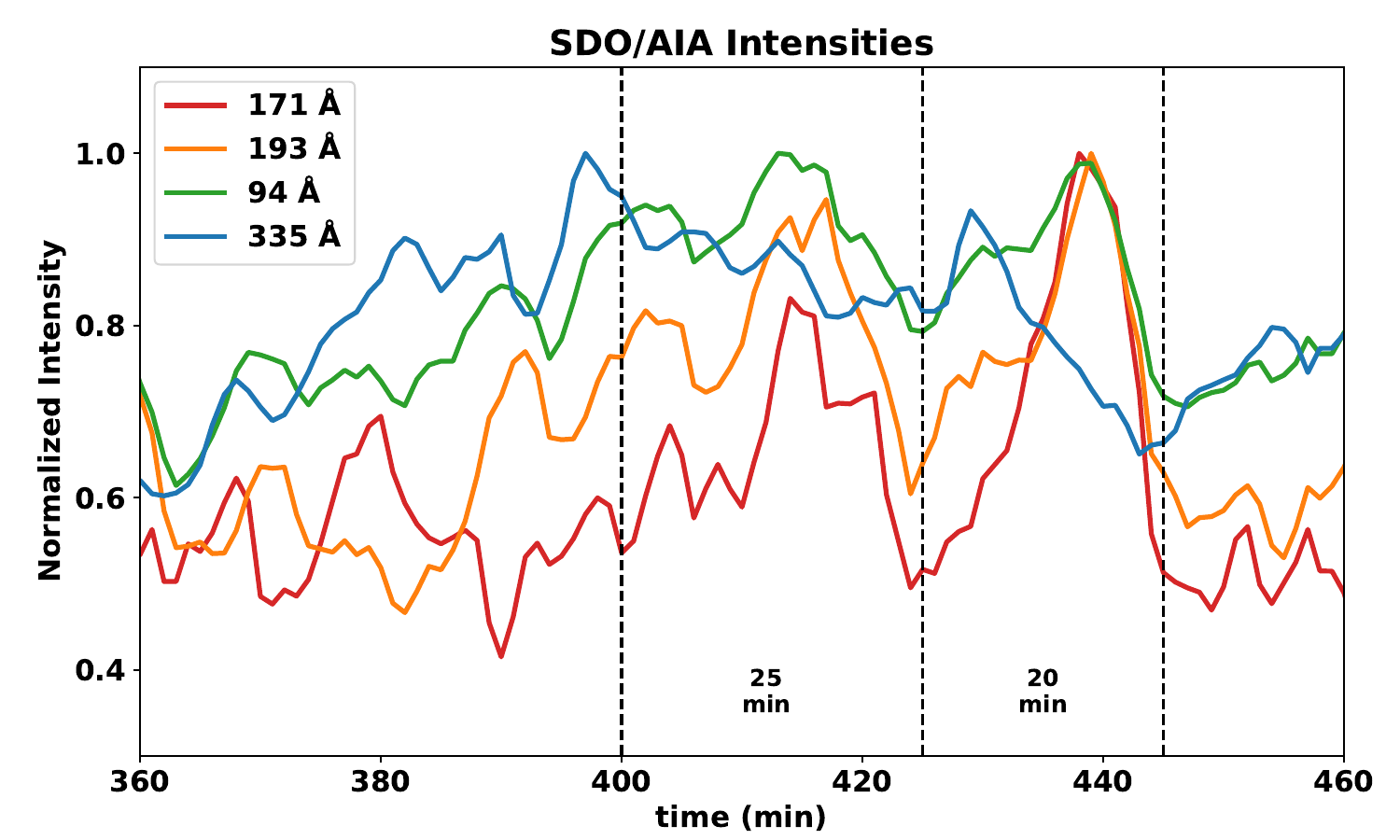}
    \vspace{-1mm}
    \caption{EUV emission characteristics of the multi-thermal clusters. 
    The plot shows the normalized intensity (emission integrated along the LOS as a function of time) in the 171~{\AA} (red), 193~{\AA} (orange), 94~{\AA} (green) and 335~{\AA} (blue) channels of AIA, corresponding to an observation made from the bottom of each emissivity map in Figure \ref{fig:storm}.
    The dashed vertical lines correspond to the lifetimes of the 193~{\AA} clusters that were identified in the Figure \ref{fig:integrated_emissivity} during the same period. 
    }
  \label{fig:storm_intensity} 
\end{figure*}
\begin{figure*}
    %
    %
    %\hspace{-4mm}
    \centering
    \includegraphics[width=0.45\textwidth]{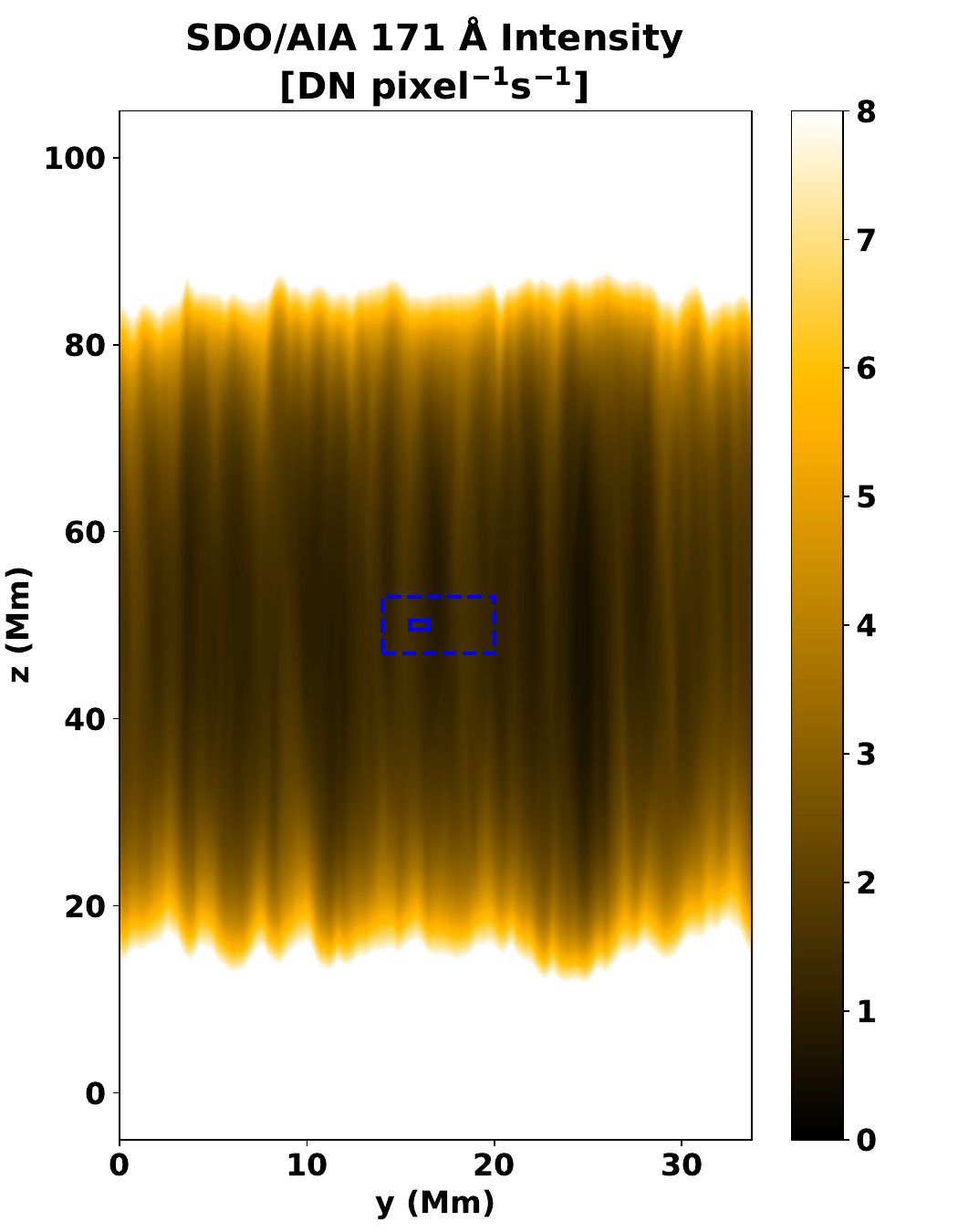}
    \includegraphics[width=0.45\textwidth]{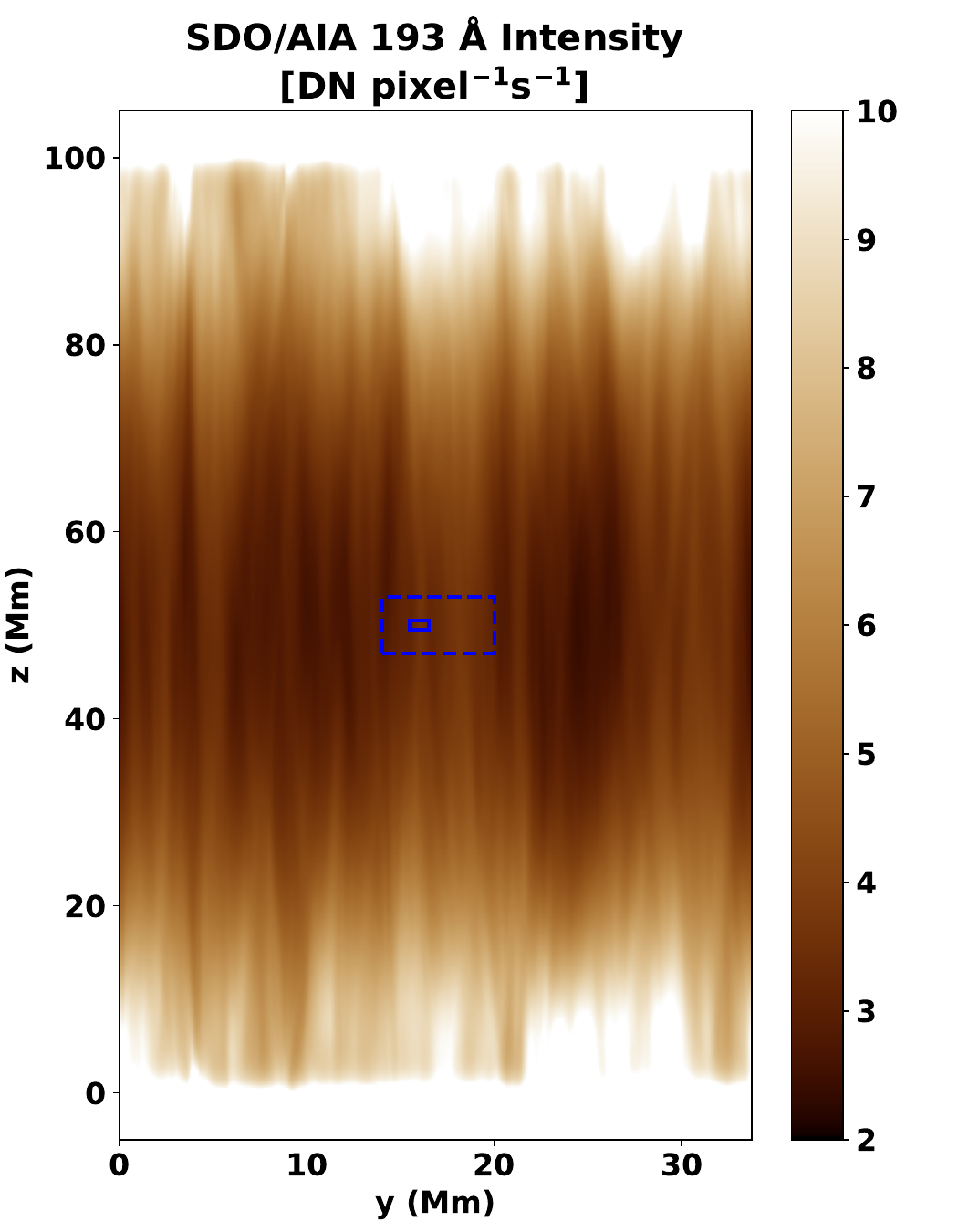}
    %
    %
    %\vspace{-5mm}
    \vspace{-2mm}
    \caption{Intensity that would be detected in the 171~{\AA} and 193~{\AA} channels of AIA at $t=440$~min, for a view from the side that corresponds to a LOS integration along $X$.
    The blue boxes mark the point where the intensity is sampled for the light curves plotted in Figure \ref{fig:storm_intensity}, while the dashed blue boxes shows a larger area spanned by the nanoflare storm.}
  \label{fig:193_image} 
\end{figure*}
\begin{figure*}
    \centering
    \includegraphics[width=0.97\textwidth]{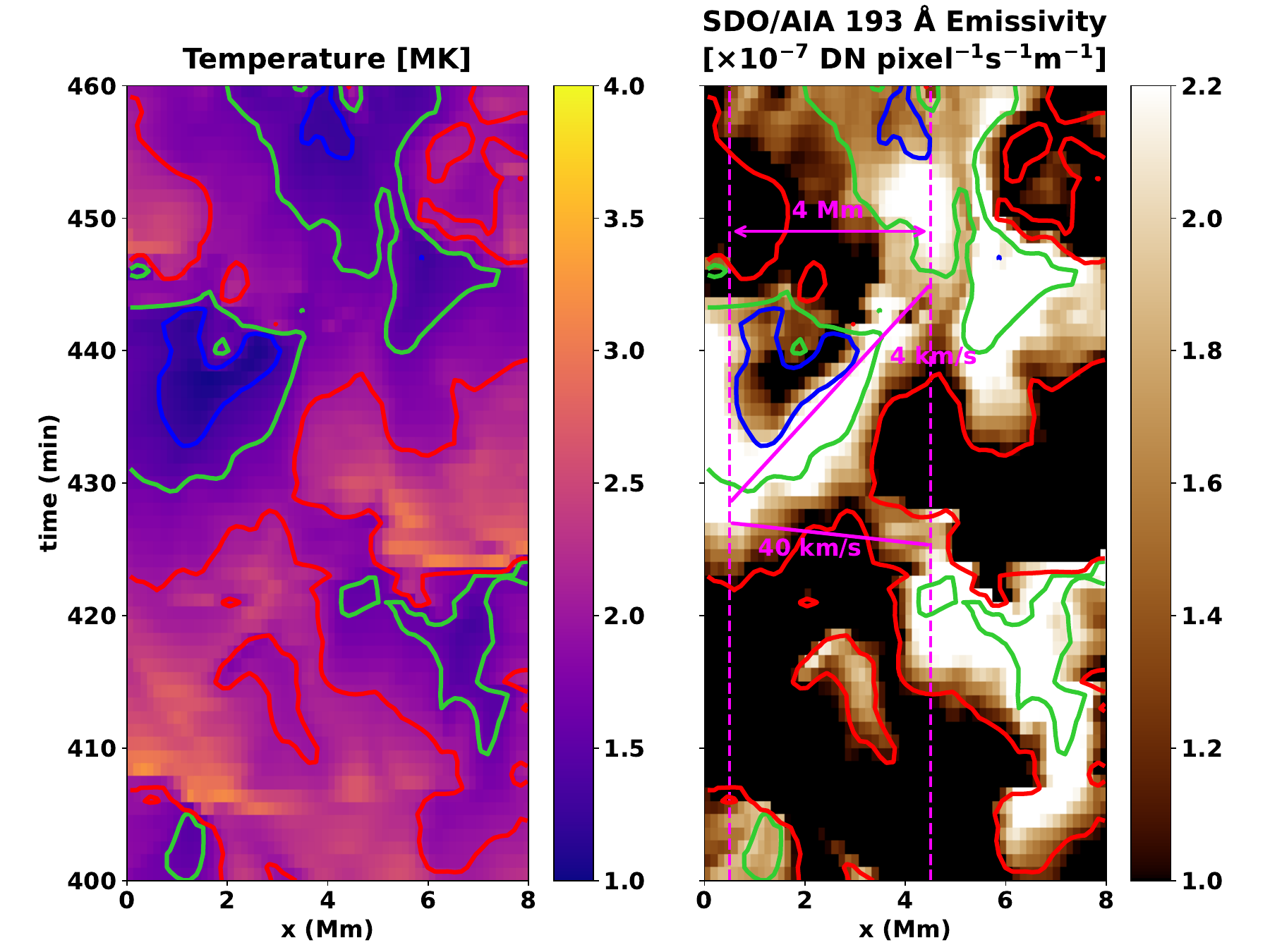}
    \vspace{-2mm}
    \caption{Close-up of Figure \ref{fig:storm} showing the transverse \lq\lq shifting\rq\rq\ of the 193~{\AA} emission, which is caused by phase differences in heating and cooling events across neighboring magnetic flux strands.
    The green contours show plasma at the peak sensitivity of the 193~{\AA} channel (1.6~MK), while the red (blue) contours indicate hotter (cooler) plasma at 2~MK (1.2~MK). 
    The pink lines on the 193~{\AA} panel mark the \lq\lq apparent\rq\rq\ displacements of the multi-thermal, multi-stranded clusters, and the slopes of these lines indicate the speeds associated with these displacements.
    }
    \label{fig:193_shifts} 
\end{figure*}
Many nanoflare storms and the subsequent clusters they produce can be identified in the movie associated with Figure \ref{fig:emissivity_midplane_xy_clusters}.
These storms all share three main characteristics. 
The first is a pronounced activity in the $J_z$ current sheets that facilitate the cascading reconnection events.
The magnetic field adjusts in response to the changing magnetic connectivity and associated changing Lorentz forces that are brought about by these reconnection events. 
This occurs not only in the immediate vicinity of the reconnection site, but all along the reconnected magnetic flux strands. 
Such dynamic behavior is especially prevalent in nanoflare storms, together with the thinning of current sheets that have not yet reconnected. 
This thinning may lead to the eventual reconnection of those sheets \citep[e.g.,][]{Rappazzo2013,Leake2024}, which is a possible mechanism for the avalanche propagation.
Other possibilities include loss of current sheet equilibrium \citep{Klimchuk2023b} and creation of a critical amount of shear across unreconnected sheets \citep{Leake2020}. 
A detailed investigation of the avalanche physics is planned for a future study. 

The second characteristic of the storms is a collective occurrence of high temperature features that are direct manifestations of the correlated nanoflares.
These high temperature features occur in quick succession over the localized area spanned by a particular storm.
The temperature decreases rapidly from its peak when the nanoflare storm ends, especially in the initial cooling phase dominated by thermal conduction. 
Density peaks later, locally in the same area as the high temperature features, after evaporation has occurred.
However, the time scale for density variation is much longer than the temperature.
Therefore, the third characteristic of the storms is a persistence of localized high density features that are formed by  the evaporative response of the plasma to the nanoflares.

Figure \ref{fig:storm} demonstrates these three characteristics for the nanoflare storm centered about the point $(x,y) = (6, 16)$~Mm, occurring between $t=360-395$~min. 
The dashed triangles on the $J_z$ current and temperature time-distance maps mark the storm and highlight the radial spread of the associated avalanche.
We also show the subsequent formation of multi-thermal clusters in the 171~{\AA}, 193~{\AA}, 94~{\AA} and 335~{\AA} channels of AIA in the lower four panels of Figure \ref{fig:storm}.
During this particular storm, the $J_z$ time-distance map shows the development of a lot of small scale structures that are co-spatial and co-temporal with the high temperature features prominent in the temperature map.
However, this hot directly heated plasma only reveals very thin faint signatures in the 94~{\AA} emissivity map, which are visible for a short period of time around $t=385$~min, within the region indicated by the red arrow.
Otherwise, as discussed above, the emission in the 94~{\AA} channel is dominated by cooler evaporated plasma.

Figure~\ref{fig:storm_intensity} shows the time evolution of the AIA intensities for LOS integrations made through each of the emissivity maps in Figure~\ref{fig:storm}.
This corresponds to a temporal sampling of the intensities at the point $y=16$~Mm, in the $Z$-midplane, as indicated by the blue boxes on the 171~{\AA} and 193~{\AA} images shown in Figure~\ref{fig:193_image}.
However, intensities averaged over a larger area spanned by the storm also show the same basic signatures.
We correlate these signatures with the evolution of the multi-thermal clusters that are formed by the storm.

The nanoflare storm ends at about $t=395$~min, and with the cessation of heating events the temperature starts to cool.
Meanwhile, evaporative upflows continue to increase the coronal density, which first peaks around $t=405$~min.
These coupled thermodynamic processes lead to the formation of the first localized cluster of brightenings seen in the 335~{\AA} and 94~{\AA} emissivity maps in Figure~\ref{fig:storm}, which are co-temporal with peaks in their respective LOS intensities.
The magnetic flux strands that comprise this cluster are then subjected to heating and cooling events at different times.
Figure~\ref{fig:193_shifts} shows a close-up of these strands to demonstrate that this forces their subsequent thermodynamic evolution to become out of phase.
In particular, the strands on the right-hand side of the cluster start to drain as they cool into the temperature range of the 193~{\AA} channel, while heating events on the left maintain the strands in that part of the cluster at temperatures above the 193~{\AA} channel, in the range of the 335~{\AA} channel.
The 193~{\AA} intensity peaks at $t=415$~min, at which time the cluster shows clear multi-thermal and multi-stranded signatures in the emissivity maps, with the out of phase magnetic flux strands producing simultaneous emission in the 335~{\AA}, 94~{\AA} and 193~{\AA} channels.

Between $t=420-425$~min, the strands on the right-hand side of the
cluster are heated by a \lq\lq substorm\rq\rq\ that increases their temperature to above the range of the 193~{\AA} channel. 
Thus, the 193~{\AA} emission plummets during this period.
However, soon thereafter, the strands on the left-hand side of the cluster finally cool into the 193~{\AA} channel.
The outcome shown in Figure \ref{fig:193_shifts} is that the 193~{\AA} part of the cluster appears to shift a transverse distance of about $4$~Mm, from right to left, in just under 2 minutes.  
This apparent rapid shifting of the cluster resembles the loop splitting evolution reported by \cite{Mandal2024,Mandal2025}, which in our simulation is caused by phase differences in heating and cooling events across neighboring magnetic flux strands.
Furthermore, the displacement identified in Figure \ref{fig:193_shifts} and the speed associated with this displacement ($\approx 40$~km/s) both show good agreement with the observationally inferred values \citep{Mandal2024,Mandal2025}.

Following the \lq\lq substorm\rq\rq\ on the right-hand side of the cluster, the density then peaks for a second time around $t=430$~min.
This effectively extends the lifetime of the first cluster in the 335~{\AA} channel so that it persists for about 60~minutes in total, between $t=375-435$~min.
In contrast, consistent with Figure \ref{fig:integrated_emissivity}, there is a clear break in the emission from the 193~{\AA} cluster. 
Therefore,
when these strands on the right then cool from the 335~{\AA} channel back into the 193~{\AA} channel, they form part of a new multi-stranded cluster in the emissivity map that is accompanied by a clear second rise phase in the 193~{\AA} intensity.

For the subsequent evolution, strands that comprise the 193~{\AA} cluster then drain and cool with a phasing that closely corresponds to that of their earlier heating events.
In particular, the magnetic flux strands on the right-hand side of the cluster experienced heating events at progressively later times than those on the left.
Thus, the strands on the left proceed to cool below the 193~{\AA} channel, while the strands on the right remain within the 193~{\AA} temperature range. 
Consequently, the 193~{\AA} emission now appears to shift $4$~Mm from the left-hand side back to the center, in just over 16 minutes, as highlighted in Figure \ref{fig:193_shifts}.
This apparent slower shifting of the cluster ($\approx 4$~km/s) shows similarities to the loop drifting evolution described by \cite{Mandal2024,Mandal2025}, with the slower \lq\lq drift\rq\rq\ speed in the simulation providing a physical explanation for the observed apparent velocities.

During this period there is also a short window when the 193~{\AA} emission is produced by two spatially separate clusters, one at the shifted cluster position (at the center), and the other at the position of the cluster before the shift (on the left-hand side).  
This type of evolution was also observed by \cite{Mandal2024,Mandal2025}.
Here, the persistence of the unshifted cluster is explained by the presence of stronger heating events on the corresponding strands, which then result in higher densities that extend the draining and cooling times beyond those on the strands that participate in the \lq\lq shifting\rq\rq .

Some of the strands that comprised the 193~{\AA} clusters continue to cool into the temperature range of the 94~{\AA} and 171~{\AA} channels, while others remain in the 193~{\AA} channel. 
The outcome shown in Figure \ref{fig:storm_intensity} is that the 193~{\AA}, 94~{\AA} and 171~{\AA} intensities all follow a very rapid evolution between $t = 437-442$~min, with their light curves peaking simultaneously at $t=440$~min.
This evolution occurs on a time scale of approximately 5 minutes compared to the much longer time scale of around 20 minutes that is typically reported for plasma cooling in coronal loops \citep{Viall2012}. 
We agree with \cite{Chitta2022} that such rapid evolution in the intensity light curves is a signature of multi-thermal and multi-stranded plasma along the LOS. 
However, our simulation highlights that this impulsive evolution in the light curves can be produced by the simultaneous cooling of plasma on different strands along the LOS, declining through different temperature ranges. Thus, it cannot be used as a unique identifier of the heating phase of impulsive events.
Instead, adopting the careful approach of \cite{Chitta2022}, additional diagnostics are required to demonstrate the rapid heating of coronal plasma in observations. 
We note, however, that rapid cooling is typically a consequence of rapid heating, even when the rapid heating cannot be directly observed.
Lastly, from $t=440$~min onward, the density continues to drain and the emission reduces accordingly.  
%
%
%%%%%%%%%%%%%%%%%%%%%%%%%%%%%%%%%%%%%%%%%%%%%%%%%%%%%%%%%%%%%%%%%%%%%%%%%%%%%%%%%%%%%%%%%%%%%%%%
%
%
\section{Summary and Discussion} \label{sec:discuss}

We have presented a self-consistent physical model that unifies the origins of both the diffuse emission and bright coronal loops observed in solar active regions \citep{Klimchuk2009,Klimchuk2015,Klimchuk2023a}.
It improves on the \cite{Klimchuk2023a} simulation in multiple ways and gives even better agreement with observations.
Within our model, diffuse emission is formed by spatially and temporally uncorrelated nanoflares, while bright coronal loops are produced by coherent clusters of nanoflares -- nanoflare storms -- that have an avalanche nature \citep{Hood2016,Reid2018,Cozzo2023}, where a single reconnection event triggers a cascade of nanoflares across neighboring unreconnected current sheets. 

Our simulation successfully reproduces several key observed characteristics of coronal loops, including their widths, lifetimes, and circular cross sections -- the latter resulting from the isotropic radial spread of reconnection events during avalanches.
The results further support the idea that coronal loops are multi-thermal, multi-stranded structures, where the observable emission is dominated by cooler evaporated plasma, while the hot directly heated plasma remains extremely faint and short-lived \citep[see e.g.,][]{Barnes2016a,Barnes2016b}. 

We have also demonstrated that impulsive variations in intensity light curves can be produced by simultaneous cooling of plasma on different strands along the LOS, highlighting the importance of using complementary diagnostics to directly identify impulsive heating events (rather than the subsequent cooling) in current observational data \citep[see e.g.,][]{Chitta2022}.
Furthermore, our simulation explains some cross-field displacements of coronal loops \citep[e.g.,][]{Mandal2024, Mandal2025} as apparent motions caused by out-of-phase heating and cooling events on neighboring magnetic flux strands.
Together, these findings support a unified picture for understanding the complex structure and thermal evolution of coronal loops, providing further insights into the heating of the magnetically closed solar corona.

Previously, \cite{Hood2016} demonstrated how an MHD avalanche can occur in a multi-stranded coronal loop.
Using a 3D MHD simulation, they showed that the kink instability of a single magnetic flux strand can rapidly propagate to neighboring stable strands, destabilizing them and triggering the release of magnetic energy in discrete bursts through magnetic reconnection. 
As this destabilization process cascades, the avalanche of instabilities produces a significant number of impulsive heating events across a large spatial scale.
Using similar models, \cite{Reid2018} went on to show that continuous driving can induce the kink instability in a single strand and subsequently destabilize neighboring strands that are also subjected to driving, while \cite{Cozzo2023} demonstrated that this avalanche process can heat the coronal plasma to temperatures of up to $10^7$~K.
However, all three models rely on initializing and/or generating a highly stressed, non-potential magnetic field with a specific structure in order to induce the kink instability and subsequent destabilizations.
After the first kink instability and avalanche occurs, the magnetic field then becomes sufficiently tangled, preventing the initiation of any further avalanches of the same type, even under continuous, coherent driving conditions.

In contrast, the avalanches that form coronal loops in our simulation are independent of the kink instability.
Instead, a single reconnection event triggers a cascade of reconnection events by modifying the physical conditions of neighboring unreconnected current sheets through the restructuring of the magnetic field and resulting changes in Lorentz forces.
This type of avalanche occurs repeatedly at different times and spatial locations throughout our multi-stand simulation, triggered whenever critical conditions are met, and is driven by the self-consistent evolution of the magnetic field in response to the imposed random photospheric motions.

Recently, \cite{Malanushenko2022} analyzed an MHD simulation of an entire active region and found that many coronal loops in their synthetic images originated from large, warped, veil-like emissivity structures in the 3D volume, rather than thin tube-like structures with circular cross sections.
Coronal loops were observed only when the veils were viewed edge-on, and not face-on.
In contrast, the loops formed in our simulation are roughly circular in shape and result from localized clusters of small-scale brightenings.
Each individual brightening can be interpreted as a small-scale, veil-like emissivity structure in 3D, with a diameter on the order of 0.5~Mm. 
These small-scale veils directly reflect the small-scale impulsive heating processes in our simulation.
In comparison, the veils in \cite{Malanushenko2022} are typically much larger, with spatial scales on the order of 10~Mm. 
This may suggest that the heating processes forming the veils in their simulation have a spatial scale of that particular size.
We note that their simulation is of an entire active region and therefore does not include the enormous number of small-scale current sheets that are present in the real corona.

In two recent studies, \cite{Mandal2024,Mandal2025} have reported the occurrence of anomalous cross-field motions in coronal loops. 
By analyzing high-resolution observations from two different vantage points -- provided by the Extreme Ultraviolet Imager \citep[EUI;][]{Rochus2020} onboard Solar Orbiter \citep{Muller2020} and the AIA onboard SDO -- they presented several examples of coronal loops exhibiting a split-drift phenomena.
In particular, the loops underwent a rapid cross-field splitting at approximately $30$~km/s, followed by a slower drifting evolution at around $5$~km/s, directed either away from or back towards the original loops.

Multiple mechanisms were proposed by \cite{Mandal2024,Mandal2025} to explain these split-drift events, including component-wise magnetic reconnection facilitating the perpendicular transport of magnetic flux strands via reconnection-driven outflow jets \citep{Pagano2021}.
Under this interpretation, the observed loop motions correspond to physical displacements of the magnetic flux strands.
However, none of the events analyzed by \cite{Mandal2024,Mandal2025}  exhibited jet-like outflows -- nanojets -- associated with the reconnection \citep{Antolin2021,Cozzo2025}, or rapid brightenings associated with the relaxation of braided magnetic fields \citep{Chitta2022}.

Therefore, our multi-stand simulation supports the more likely scenario in which the observed cross-field displacements of the coronal loops correspond to apparent rather than physical motions.
Specifically, we demonstrated that phase differences in the thermal evolution of neighboring magnetic flux strands provide a physical explanation for the apparent rapid cross-field splitting and slower drifting motions observed by \cite{Mandal2024,Mandal2025}. 
The apparent displacements and associated velocities produced in our simulation show good agreement with the observationally inferred values.
While reconnection fundamentally drives the heating and subsequent thermodynamic evolution of each individual magnetic flux strand, the apparent motions arise from heating and cooling events that are out-of-phase across neighboring strands.
Thus, since these processes are confined to field-aligned dynamics, they do not produce any real cross-field motions.

Lastly, we note that the magnetic flux on the Sun is concentrated in clumps in the photosphere and expands with height into the corona, while our simulation has considered only the case of a straight, uniform magnetic field.
Thus, the magnetic structure in our simulation artifically allows nanoflares to occur uniformly at all heights with equal distributions of energy release. 
Consequently, coronal nanoflares control the plasma response due to the lower densities in which they occur, relative to chromospheric events (see Daldorff et. al, in preparation).
However, simulations that include realistic, non-uniform flux distributions and consider the expansion of the magnetic field with height show a stratification of heating that scales with the magnetic field strength \citep{Downs2016,Kohutova2020,Antolin2022b}. 
Therefore, in future studies, we plan to improve our model by incorporating these two important effects.
This may lead to a non-uniform height distribution of nanoflares that facilitates the occurrence of thermal non-equilibrium \citep{Antiochos2000,Karpen2001,Antolin2010,Mikic2013,Froment2018,Johnston2019b,Klimchuk2019,Johnston2025}.
%
%
%%%%%%%%%%%%%%%%%%%%%%%%%%%%%%%%%%%%%%%%%%%%%%%%%%%%%%%%%%%%%%%%%%%%%%%%%%%%%%%%%%%%%%%%%%%%%%%%
%
%
\section{Acknowledgments}

CDJ, LKSD, JAK, SSM, and JEL acknowledge support from the NASA Heliophysics Internal Scientist Funding Model (H-ISFM) competitive work package program (PI Jim Klimchuk).
JR acknowledges financial support from the Science and Technology Facilities Council (STFC) through Consolidated Grant ST/W001195/1 to the University of St Andrews.
We also acknowledge useful discussions with Peter W. Schuck, Mark G. Linton, Kalman J. Knizhnik, N. Dylan Kee and Yi-Min Huang.
Simulations were performed by CDJ on the NASA High-End Computing Capability (HECC) using NASA Advanced Supercomputing (NAS). 
Source code, raw data, and plotting routines related to this publication are available upon request. 
We also thank the referee for their helpful comments that improved the manuscript.
%
%
%%%%%%%%%%%%%%%%%%%%%%%%%%%%%%%%%%%%%%%%%%%%%%%%%%%%%%%%%%%%%%%%%%%%%%%%%%%%%%%%%%%%%%%%%%%%%%%%
%
\bibliography{CDJ_bibliography}{}
\bibliographystyle{aasjournal}
%
%
%%%%%%%%%%%%%%%%%%%%%%%%%%%%%%%%%%%%%%%%%%%%%%%%%%%%%%%%%%%%%%%%%%%%%%%%%%%%%%%%%%%%%%%%%%%%%%%%
%
\end{document}